\documentclass[12pt, draftclsnofoot, onecolumn]{IEEEtran}

\ifCLASSINFOpdf

\else

\fi
\usepackage[cmex10]{amsmath}
\usepackage{amssymb}
\usepackage{cite}
\usepackage{graphicx}
\usepackage{array,color}
\usepackage{amsmath}
\usepackage{stfloats}
\usepackage{graphicx}
\usepackage{subfigure}
\usepackage{tabularx}
\usepackage{epsfig,epsf,color,balance,cite}
\usepackage{algorithmic}
\usepackage{algorithm}
\usepackage{bm}
\usepackage{textcomp}

\newtheorem{rem}{Remark}

\begin{document}

\title{Energy Efficiency Optimization for MIMO Distributed Antenna Systems}

\author{

{Hong Ren, Nan Liu, Cunhua Pan, Chunlong He}

\thanks{This work is partially supported by the National Basic Research Program of China (973 Program 2012CB316004), the National Natural Science Foundation of China under Grants $61571123$, $61201170$ and $61221002$, and Qing Lan Project.}

\thanks{Hong Ren, Nan Liu, Cunhua Pan are with National Mobile Communications Research Laboratory, Southeast University, Nanjing 210096, China. (Email:\{renhong, nanliu and cunhuapan\}@seu.edu.cn).}

\thanks{Chunlong He is with the College of Information Engineering, Shenzhen, Shenzhen University, 518060, China (E-mail: chunlonghe@163.com).}

}

\maketitle\linespread{1.6}

\begin{abstract}
In this paper, we propose a transmit covariance optimization method to maximize the energy efficiency (EE) for a single-user distributed antenna system, where both the  remote access units (RAUs) and the user are equipped with multiple antennas. Unlike previous related works, both the rate requirement and RAU selection are taken into consideration. Here, the total circuit power consumption is related to the number of active RAUs. Given this setup, we first propose an optimal transmit covariance optimization method to solve the EE optimization problem under a fixed set of active RAUs. More specifically, we split this problem into three subproblems, i.e., the rate maximization problem, the EE maximization problem without rate constraint, and the power minimization problem, and each subproblem can be efficiently solved. Then, a novel distance-based RAU selection method is proposed to determine the optimal set of active RAUs. Simulation results show that the performance of the proposed RAU selection is almost identical to the optimal exhaustive search method with significantly reduced computational complexity, and the performance of the proposed algorithm significantly outperforms the existing EE optimization methods.\\

\end{abstract}

\begin{keywords}
 Distributed antenna system,  multiple antennas, energy efficiency, rate constraints, RAU selection
\end{keywords}

\IEEEpeerreviewmaketitle

\section{Introduction}

Recently, distributed antenna system (DAS) has been regarded as a promising technique to meet the increasing demand for high spectral efficiency (SE, bit/s/Hz) and good coverage \cite{xiaohuyou,Huiling2011,jiangzhou2012}. Unlike the conventional centralized antenna system (CAS),  several distributed RAUs with one or more antennas  are geographically separated over the cell in DAS. Each RAU is connected to the central processing unit via a high-speed error-free low-latency wired channel such as optical fiber links. Through this distributed implementation, the average access distance for each user can be significantly reduced. Thus, the DAS has the great potential to improve the SE and coverage, especially for the cell-edge users. On the other hand, large-scale multiple-input and multiple-output (MIMO) has attracted extensive interests due to its  advantage of providing tremendous SE for wireless communications \cite{Marzetta}. Placing a large number of antennas in different locations can also reduce the correlations of antennas \cite{Sun2013}, thus improving the SE of the system.

The SE of the DAS system has been extensively studied in the past few years \cite{Choi2007,Vu2011,Sang2013,Heath2011}. Specifically, \cite{Choi2007} derived the downlink capacity of a DAS under two cases: unknown channel state information (CSI) and know CSI. In the considered DAS, both the RAUs and the users are equipped with only one antenna. For the case of known CSI, the authors provided one optimal solution for the transmit covariance matrices. However, the property of the optimal solution was not characterized. Then, in \cite{Vu2011}, the authors solved a relaxed problem by replacing the original positive semi-definite constraint with a $2 \times 2$ matrix minor condition, and showed that the optimal phases of the beam weights at the RAUs should be matched to the phases of the channel coefficients, and the amplitudes are determined by the power constraints. Similar method and result can be found in \cite{Sang2013} for a DAS system with multiple antennas at each RAU and one antenna for each user. The optimal beam direction for each RAU should be matched to its channel vector and each RAU should use full power to transmit. In \cite{ruizhang2010}, the author considered the precoding design to maximize the sum-rate for a multiuser MIMO DAS system. Block diagonalization method was applied to eliminate the multiuser interference and the system is reduced to a multiple parallel single-user MIMO DAS system. However, the rate requirement was not considered in the paper.

On the other hand, energy efficiency (EE) has attracted extensive interests \cite{Daquan2013,panwcletter2015} due to energy shortage and the greenhouse effect, and it will be one of the main concerns for the fifth generations (5G) mobile networks \cite{Andrews-2014}. The EE is defined as the ratio of the rate to the total power consumption, representing the number of bits that can be successfully transmitted with per Joule energy consumption. In general, the SE maximization solution for the DAS may not be the EE maximization solution of the DAS since the former always activates all RAUs and uses full power to transmit. Hence, there have been some literatures studying the EE maximization problem for the DAS \cite{Heejin2013,chen2012energy,Heejincletter,Chunlong2014globsip,Onireti2013}. In \cite{Heejin2013}, the authors designed a power allocation method for single-user DAS systems. By solving the Karush-Kuhn-Tucker conditions, the optimal solution can be obtained in closed form. The EE power allocation for the frequency-selective fading DAS was studied in \cite{chen2012energy}, where a numerical search method was proposed. \cite{Heejincletter} extended the work of \cite{Heejin2013} to a more general DAS where each RAU has multiple antennas by considering both beamforming and power allocation. The authors showed that the optimal beam direction for the EE maximization should be matched to the channel vector as in \cite{Sang2013}, and the optimal power allocation can be obtained in closed form by the same method in \cite{Heejin2013}. However, all of these works do not consider the user's quality of service such as the rate requirements, and thus the attained rate could be fairly low. In addition, RAU selection or antenna selection are not taken into account in these papers.  When the user's rate requirement can be satisfied, shutting off some unnecessary RAUs can save circuit power consumption, and thus improves the EE of DAS. Also, the above papers focused on the DAS where the user is equipped with only one antenna. Due to the fast development in antenna technology \cite{Sai2014}, multiple antennas may be packed in each user's terminal. In this case, the multiple-antenna RAUs can transmit multiple streams to the user to enhance the SE of the DAS.  To the best of our knowledge, there is few works studying the EE maximization problem for this more general single-user MIMO DAS \cite{Chunlong2014globsip,Onireti2013}. In  \cite{Chunlong2014globsip}, the authors obtained an approximate EE expression for this general scenario based on the assumption of the high SNR approximation. Then, a more tight closed-form approximation of the EE-SE trade-off  over the Rayleigh fading channel was derived in \cite{Onireti2013}.  However, the above works studied the EE of the  MIMO DAS from the performance analysis point of view and did not provide any insights on how to design the optimal transmit strategy. In the MIMO DAS, the beam directions and power allocation over the beams should be jointly optimized, and the beam directions cannot be obtained in the closed form as in the MISO case \cite{Heejincletter}. Some iterative methods should be developed to find the optimal beam direction.

In this paper, we study the EE maximization problem for the single-user downlink MIMO DAS, where both the RAUs and the user are equipped with multiple antennas. Both the rate requirement and the RAU selection are taken into consideration. We propose an iterative optimization method to alternatively optimize the transmit covariance matrices and the set of active RAUs.

Under given active RAUs, the optimization problem is a pseudo-concave maximization problem, which can be solved by the standard convex optimization techniques such as the interior point method \cite{p6-boyd2004convex}. However, this method will incur heavy computational complexity and cannot reveal the structure of the optimal transmit covariance matrix. Hence, the approach we take in this paper is to divide the optimization problem into three subproblem, i.e., the rate maximization problem, the EE maximization problem without rate constraint and the power minimization problem. For the rate maximization problem with per-RAU power constraint, we devise an efficient algorithm by resorting to the dual-decomposition based method and the closed-form expression of the transmit covariance matrix can be derived. If the obtained maximum rate satisfies the rate requirement, we then attempt to solve the EE maximization problem without rate constraint, which can be easily solved by fractional programming. If the obtained solution satisfies the rate requirement, it is the optimal solution of the optimization problem given the set of active RAUs. Otherwise, we need to solve the power minimization problem with the rate constraint, which can be effectively solved by the bisection search method. In addition, we analyze the complexity to solve each subproblem, and show that our proposed algorithm yields a much lower complexity than the standard convex optimization techniques.

Next, a novel distance-based RAU selection procedure is proposed to determine the optimal set of active RAUs. This selection method is reasonable since  user's channel gain is mainly determined by the distance, and in general there are great differences among the distances from the user to the RAUs  of the DAS. This method has a lower complexity compared with the channel-norm based antenna selection method in \cite{Jie2013}, as the calculation of channel-norms incurs additional computation complexity. Moreover, the search procedure will stop once the current achieved EE is lower than the previous highest EE, and the set of active RAUs achieving the previous highest EE will be the output of our algorithm. Simulation results show that the proposed distance-based RAU selection and transmit covariance optimization scheme yields almost the same EE performance as the exhaustive RAU selection, and significantly outperforms the existing EE optimization method.

The rest of this paper is organized as follows. In Section \ref{systemmodel}, we introduce the system model and problem formulation. Section \ref{EEfixedsec} develops the optimal transmit covariance optimization method to deal with the EE maximization problem under a given set of active RAUs. Section \ref{portselection} presents the novel low-complexity distance-based RAU selection  method. Simulation results are given in Section \ref{simulation} to evaluate the performance of the proposed RAU selection method. Finally, Section \ref{conclusion} concludes this paper.

\section{System Model And Problem Formulation}\label{systemmodel}

\begin{figure}
\centering
\includegraphics[width=3.5in]{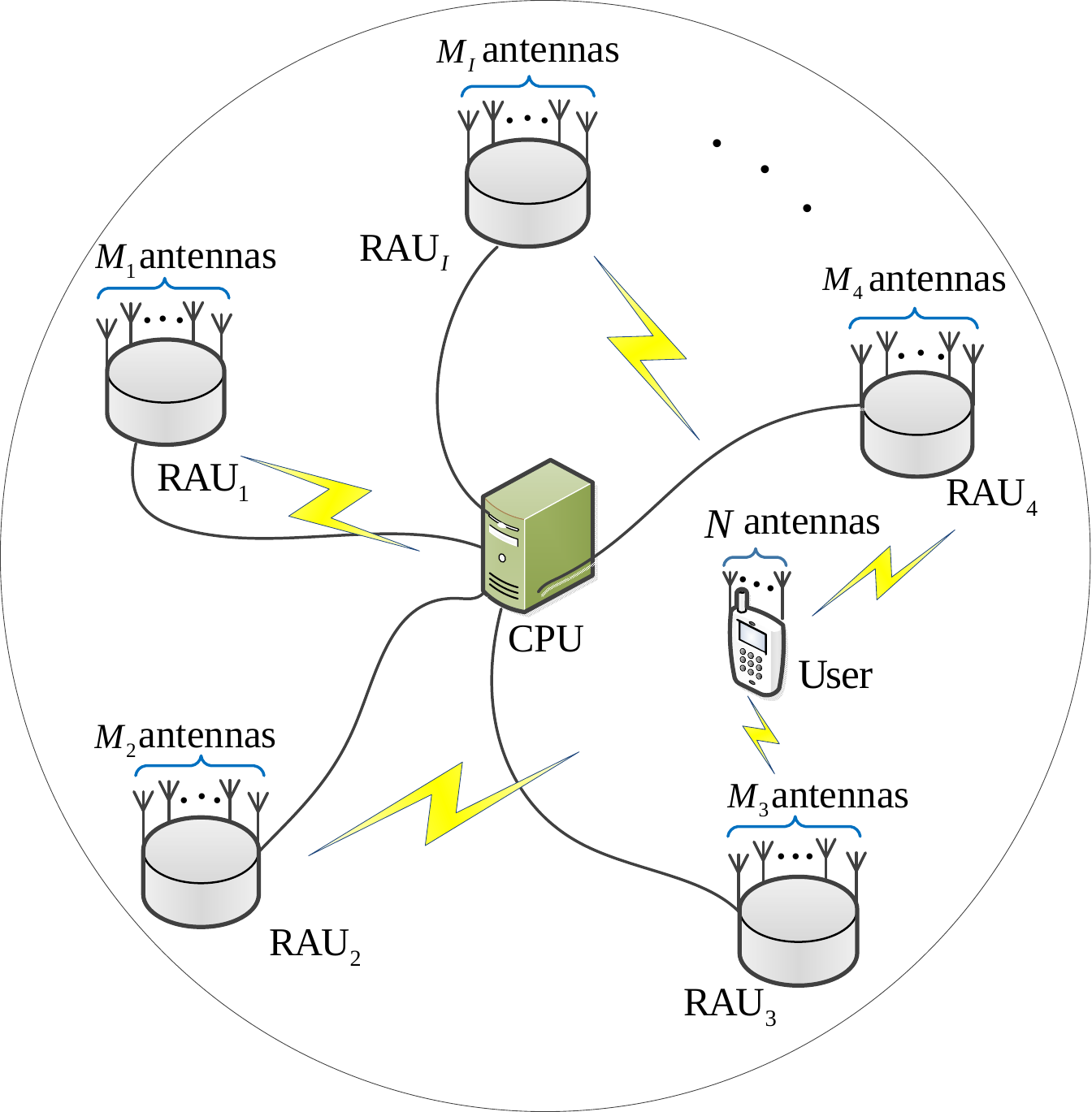}
\caption{Structure of DAS with $I$ distributed antenna RAUs.}
\label{fig1}
\end{figure}

We consider a downlink single cell DAS with $I$ RAUs and one user, where RAU $i$ is equipped with ${M_i}$  antennas and the user is equipped with $N$ antennas, $i=1,2,\ldots,I$, as shown in Fig. \ref{fig1}. We assume that all the RAUs are connected to the central processing unit (CPU) through the high speed fiber-optic cable. Moreover, it is assumed that all the RAUs are fully controlled by the CPU.

If we activate all RAUs, the system may suffer EE performance loss due to the circuit power consumption of all RAUs. Hence, selecting a subset of RAUs may be a good option in terms of the EE performance.  Denote $\mathbb{S} \subseteq \left\{ {1,2, \cdots ,I} \right\}$ as the set of the selected RAUs to transmit signals to the user with the number of RAUs $A \triangleq \left| \mathbb{S} \right|$. The subset $\mathbb{S}$ is given as $\mathbb{S}= \left\{ {{s_1},{s_2}, \cdots ,{s_A}} \right\}$, whose elements represent the RAU indices and are arranged in increasing order, i.e., ${s_1} < {s_2} <  \cdots  < {s_A}$. When the selected active RAU set is $\mathbb{S}$, the downlink transmission can be modeled as a single user MIMO with ${M_A}$ transmitting antennas and $N$ receiving antennas, where ${M_A} = \sum\nolimits_{i = 1}^{A} {{M_{{s_i}}}} $ denotes the total number of active antennas.

Let ${{\bf{H}}_{{s_i}}} \in {\mathbb{C}^{N \times {M_{{s_i}}}}},{s_i} \in \mathbb{S}$ be the channel matrix from RAU ${s_i}$ to the user, and ${{\bf{H}}_ \mathbb{S} } = \left[ {{{\bf{H}}_{{s_1}}},{{\bf{H}}_{{s_2}}}, \cdots ,{{\bf{H}}_{{s_A }}}} \right] \in {\mathbb{C}^{N \times {M_A}}}$ be the overall channel matrix from the selected RAUs in $\mathbb{S}$ to the user. Hence, the received signal at the user from all RAUs is given by
\begin{equation}
\label{eq1}
{{\bf{y}}_ \mathbb{S}} = {{\bf{H}}_\mathbb{S}}{{\bf{x}}_\mathbb{S}} + {\bf{z}}
\end{equation}
where ${{\bf{x}}_\mathbb{S}} \in{\mathbb{C}^{{M_A} \times 1}}$ and ${{\bf{y}}_ \mathbb{S} } \in{\mathbb{C}^{N \times 1}}$ denote the transmit and receive signals for the user, respectively, and ${\bf{z}} \in {\mathbb{C}^{N \times 1}}$ denotes the receiver noise at the user. For simplicity, we assume that ${\bf{z}} \sim {\cal C}{\cal N}({\bf{0,I}})$. It is also assumed that the CPU can acquire perfect channel state information (CSI), and the channels are in block-fading and change sufficiently slowly such that they can be treated as fixed during the considered transmission period.

 Denote ${{\bf{Q}}_\mathbb{S}} =\mathbb{E} {\rm{\{ }}{{\bf{x}}_\mathbb{S}}{\bf{x}}_\mathbb{S}^H{\rm{\} }}$ as the transmit covariance matrix for the user, with ${{\bf{Q}}_{ \mathbb{S} }} \in {\mathbb{C}^{{M_A} \times {M_A}}}$ and ${{\bf{Q}}_{ \mathbb{S} }}\succeq{\bf{0}}$. Then, the spectral efficiency (SE) (bps/Hz) for the user is \cite{Telatar1999}
\begin{equation}
\label{eq2}
{R_{ \mathbb{S} }} = {\log _2}\left| {{\bf{I}} + {{\bf{H}}_{ \mathbb{S} }}{{\bf{Q}}_{\mathbb{S} }}{\bf{H}}_{\mathbb{S} }^H} \right|.
\end{equation}

In order to impose the per-RAU power constraint, we introduce a set of matrices, ${{\bf{B}}_i},i = 1, \cdots ,A$, as follows
\begin{equation}
\label{eq4}
{{\bf{B}}_i} \buildrel \Delta \over = {\rm{Diag}}\left( {\underbrace {0, \cdots ,0,}_{\sum\nolimits_{l = 1}^{i - 1} {{M_{{s_l}}}} }\underbrace {1, \cdots ,1,}_{{M_{{s_i}}}}\underbrace {0, \cdots ,0}_{\sum\nolimits_{l = i + 1}^A {{M_{{s_l}}}} }} \right),i = 1, \cdots ,A.
\end{equation}
Then, the per-RAU power constraint can be expressed as
\begin{equation}
\label{eq3}
{\rm{tr(}}{{\bf{B}}_i}{{\bf{Q}}_{ \mathbb{S} }}{\rm{)}} \le {P_{{s_i}}},{\rm{  }}i = 1, \cdots ,A,
\end{equation}
where ${P_{{s_i}}}$ denotes the transmit power constraint for RAU ${s_i}$.

To consider the energy-efficiency design, the total power consumption should be considered: the power for reliable data transmission and the circuit power consumption, which is the power consumed by mixers, filters and digital-to-analog converters, digital signal processing (DSP), RF chains, etc. Hence, the total power consumption can be modeled as \cite{p11-he2013coordinated}
\begin{equation}
\label{eq5}
{P_{{\rm{total}}}} = {{\rm{tr}}\left( {{{\bf{Q}}_{\mathbb{S}}}} \right)} + M_A{p_c}+Ap_0,
\end{equation}
where  ${p_c}$ denotes the RF chain power consumption corresponding to one antenna, and $p_0$ is the static power consumption for each RAU. For simplicity, define $P_{C,\mathbb{S}} \triangleq  M_A{p_c}+Ap_0$.

The EE (bits/Joule) is defined as the ratio of the the number of bits to the total power consumption. Our objective is to find a selected active RAU set ${ \mathbb{S} } \subseteq \{{1,2, \cdots ,I}\}$ and optimize the corresponding transmit covariance matrix ${{\bf{Q}}_{ \mathbb{S} }}$ to maximize the EE of the DAS system, subject to both the per-RAU power constraint and the rate requirements. Formally, this problem can be formulated as
\begin{equation}
\label{eq7}
\begin{array}{l}
{\rm{(\textbf{P0})}}\ \mathop {{\rm{max}}}\limits_{\mathbb{S},{{\bf{Q}}_\mathbb{S}} \succeq {\bf{0}}} \ \frac{W{{{\log }_2}\left| {{\bf{I}} + {{\bf{H}}_\mathbb{S}}{{\bf{Q}}_\mathbb{S}}{\bf{H}}_\mathbb{S}^H} \right|}}{{{{\rm{tr}}\left( {{{\bf{Q}}_\mathbb{S}}} \right)} + P_{C,\mathbb{S}}}}\\
\qquad\quad\; {\rm{s.t.}}\quad\; {\rm{ tr}}\left( {{{\bf{B}}_i}{{\bf{Q}}_\mathbb{S}}} \right) \le {P_{{s_i}}},\;\;\forall i = 1, \cdots ,A,\\
\qquad\qquad\quad\;\;\; W{\log _2}\left| {{\bf{I}} + {{\bf{H}}_\mathbb{S}}{{\bf{Q}}_\mathbb{S}}{\bf{H}}_\mathbb{S}^H} \right| \ge {R_{\min }},
\end{array}
\end{equation}
where $W$ is the channel bandwidth and ${R_{{\rm{min}}}}$ denotes the minimum rate requirements.

For Problem $(\rm{\textbf{P0}})$, it is difficult to jointly optimize the selected RAU set $\mathbb{S}$ and the transmit convariance matrix ${{\bf{Q}}_{ \mathbb{S} }}$ since both the channel matrices and the total power consumption depend on $\mathbb{S}$. Fortunately, given $\mathbb{S}$, the numerator and denominator of the objective function in (\ref{eq7}) are concave and affine respectively, and both of them are differential. Hence, the objective function in (\ref{eq7}) is pseudo-concave. Moreover,  the feasible set in Problem $(\rm{\textbf{P0}})$ is convex. Thus, the optimization Problem $(\rm{\textbf{P0}})$ under given $\mathbb{S}$ is fractional programming \cite{Isheden2012}. For this kind of problem, any local optimal solution is also the global optimal solution, which can be solved through the standard convex optimization method \cite{boyd2004convex}. In the following two sections, we first optimize the transmit convariance matrix under given $\mathbb{S}$ and then provide several techniques to determine $\mathbb{S}$, respectively.

\section{EE Optimization for Given RAU Set}\label{EEfixedsec}

In this section, we optimize the transmit convariance matrix to maximize EE under given $\mathbb{S}$. Though this problem can be solved through the standard convex optimization method such as the interior-point method \cite{boyd2004convex}, it will incur considerable computational complexity when the RAU number becomes large. Hence, an efficient algorithm with low complexity is appealing, which will be the aim of this section. For notational simplicity, we omit $\mathbb{S}$ in the subscript in this section. Define ${\cal W}$ as
\begin{equation}
\begin{array}{l}
\label{eq25}
{\cal W} = \left\{ {{{\bf{Q}}} \in {\mathbb{C}^{{M_A} \times {M_A}}}:{\bf{Q}}\succeq{\bf{0}},{\rm{tr}}\left( {{{\bf{B}}_i}{\bf{Q}}} \right) \le {P_{{s_i}}},\;\;\forall i } \right\}.
\end{array}
\end{equation}
Then, by discarding the constant $W$, we can rewrite the problem as follows
\begin{equation}
\label{EEwithfixedset}
\begin{array}{l}
\mathop {{\rm{max}}}\limits_{{\bf{Q}} \in {\cal W}} \ \frac{{{{\log }_2}\left| {{\bf{I}} + {\bf{H}}{\bf{Q}}{\bf{H}}^H} \right|}}{{{{\rm{tr}}\left( {{\bf{Q}}} \right)} + P_C}}\\
\;\;{\rm{s.t.}}\;\;\; {\log _2}\left| {{\bf{I}} + {\bf{H}}{\bf{Q}}{\bf{H}}^H} \right| \ge {\tilde R_{\min }}.
\end{array}
\end{equation}
where ${\tilde R_{\min }} = {{{R_{\min }}} \mathord{\left/
 {\vphantom {{{R_{\min }}} W}} \right.
 \kern-\nulldelimiterspace} W}$.

Before solving this problem, we introduce three auxiliary subproblems
\begin{eqnarray}
&(\rm{\textbf{P1}})& \ {{\bf{Q}}_{(\text{P1})}^*} = \arg \mathop {\max }\limits_{{\bf{Q}} \in {\cal W}} {\log _2}\left| {{\bf{I}} + {\bf{HQ}}{{\bf{H}}^H}} \right|,\label{P1}\\
&{\rm{(\textbf{P2})}}&\ {{\bf{Q}}_{(\text{P2})}^*} = \arg \mathop {{\rm{max}}}\limits_{{\bf{Q}} \in {\cal W}} \;\frac{{{{\log }_2}\left| {{\bf{I}} + {\bf{HQ}}{{\bf{H}}^H}} \right|}}{{{{\rm{tr}}\left( {\bf{Q}} \right)} + P_C}},\label{P2}\\
&{\rm{(\textbf{P3})}}&\ {{\bf{Q}}_{(\text{P3})}^*} \!\!=\!\! \arg \mathop {{\rm{min}}}\limits_{{\bf{Q}} \in {\cal W}, {{\log }_2}\left| {{\bf{I}} + {\bf{HQ}}{{\bf{H}}^H}} \right| = {\tilde R_{\min }}}\! \!\!{\rm{tr(}}{\bf{Q}}{\rm{)}}. \label{P3}
\end{eqnarray}

Due to the rate constraint, the problem in (\ref{EEwithfixedset}) may be infeasible as shown in \cite{cunhua2014}. Hence, we need to check the feasibility of the problem in (\ref{EEwithfixedset}), which can be accomplished by checking whether the maximum SE under all RAUs' power constraints can satisfy the rate requirement.  The rate maximization problem under all RAUs' power constraints is given in Problem $(\rm{\textbf{P1}})$. If the optimal solution to Problem $(\rm{\textbf{P1}})$, denoted  as ${\bf{Q}}_{(\text{P1})}^*$, can satisfy the rate constraint, this problem is feasible. Otherwise, it is infeasible. When this problem is feasible, we continue to solve the EE optimization problem without the rate constraint, which is given in Problem $(\rm{\textbf{P2}})$. Denote the optimal solution to Problem $(\rm{\textbf{P2}})$ as ${\bf{Q}}_{(\text{P2})}^*$. If ${\log _2}\left| {{\bf{I}} + {\bf{H}}{\bf{Q}}_{(\text{P2})}^*{{\bf{H}}^H}} \right| \ge {\tilde R_{\min }}$ holds, ${\bf{Q}}_{(\text{P2})}^*$ is the optimal solution to the problem in (\ref{EEwithfixedset}) since Problem $(\rm{\textbf{P2}})$ is a relaxed version of the problem in (\ref{EEwithfixedset}). Otherwise, the rate constraint are met with equality at the optimum. Hence, the problem in (\ref{EEwithfixedset}) reduces to the power minimization problem given in Problem $(\rm{\textbf{P3}})$.  Denote the optimal solution to Problem $(\rm{\textbf{P3}})$ as ${\bf{Q}}_{(\text{P3})}^*$, which is the optimal solution to the problem in (\ref{EEwithfixedset}).

In summary, the algorithm to solve the problem in (\ref{EEwithfixedset}) is given in Algorithm 1. The remaining task is to solve Problems $(\rm{\textbf{P1}})$, $(\rm{\textbf{P2}})$ and $(\rm{\textbf{P3}})$, respectively. The details of solving these subproblems are given in the following subsections.
\begin{algorithm}
\caption{Solving EE optimization problem under fixed $\mathbb{S}$}

\quad 1. Solve Problem $(\rm{\textbf{P1}})$ to obtain the optimal solution ${\bf{Q}}_{(\text{P1})}^*$. If ${\log _2}\left| {{\bf{I}} + {\bf{H}}{{\bf{Q}}_{(\text{P1})}^*}{{\bf{H}}^H}} \right| \ge {\tilde R_{\min }}$, go to Step 2; Otherwise, declare that the problem is infeasible and terminate.

\quad 2. Solve Problem $(\rm{\textbf{P2}})$ to obtain the optimal solution ${\bf{Q}}_{(\text{P2})}^*$. If ${\log _2}\left| {{\bf{I}} + {\bf{H}}{\bf{Q}}_{(\text{P2})}^*{{\bf{H}}^H}} \right| \ge {\tilde R_{\min }}$, the optimal solution to the problem in (\ref{EEwithfixedset}) is ${\bf{Q}}_{(\text{P2})}^*$, terminate; Otherwise, go to Step 3.

  \quad 3. Solve Problem $(\rm{\textbf{P3}})$ to obtain the optimal solution ${\bf{Q}}_{(\text{P3})}^*$, which is the optimal solution to the problem in (\ref{EEwithfixedset}).

\end{algorithm}

\subsection{Algorithm to solve Problem $(\rm{\textbf{P1}})$}

When the user is equipped with only one antenna, the sum-rate maximization problem  $(\rm{\textbf{P1}})$ has been solved recently in \cite{Sang-2013}, in which the optimal beamforming is obtained by solving a relaxed problem by replacing the positive semi-definite constraint with a $2\times2$ matrix minor condition. The authors in \cite{Sang-2013} shows that the optimal solution takes a form of maximum ratio transmission per each RAU with full power. However, the method in \cite{Sang-2013} is not applicable for the MIMO case considered here and a new method needs to be developed.

Obviously, Problem $(\rm{\textbf{P1}})$ is a convex problem and thus can be solved using standard convex optimization techniques, e.g., the interior-point method. However, such an approach has high complexity. Instead, we aim to provide a low-complexity algorithm for solving Problem $(\rm{\textbf{P1}})$ via its dual problem due to the zero gap between Problem $(\rm{\textbf{P1}})$ and its dual problem. Note that this method has been extensively used to develop efficient beam-vectors and precoding matrices \cite{Wei2007,Dahrouj2010,ruizhang2010,Rui2010tsp,cunhua2014}.

We first derive the Lagrangian function of $(\rm{\textbf{P1}})$ with respect to all RAUs' power constraints, given by
\begin{equation}
\begin{array}{l}
L({{\bf{Q}}},\{ {\lambda _i}\} ) = {\rm{lo}}{{\rm{g}}_2}\left| {{\bf{I}} + {{\bf{H}}}{{\bf{Q}}}{\bf{H}}^H} \right|{\rm{ + }}\sum\limits_{i = 1}^A {{\lambda _i}\left( {{P_i} - {\rm{tr}}\left( {{{\bf{B}}_i}{\bf{Q}}} \right)} \right)},
\end{array}
\end{equation}
where $\{ {\lambda _i} \ge 0,\forall i\} $  are the corresponding Lagrangian dual variables. Therefore, the dual function is
\begin{equation}
\begin{array}{l}
\label{eq10}
g(\{ {\lambda _i}\} ) = \mathop {\textmd{max}}\limits_{{\bf{Q}}\succeq {\bf{0}}} L({\bf{Q}},\{ {\lambda _i}\} ).
\end{array}
\end{equation}
Then the dual problem is given by
\begin{equation}
\begin{array}{l}
\label{eq11}
\mathop {\min }\limits_{\{ {\lambda _i} \ge 0,\forall i\} } g(\{ {\lambda _i}\} ).
\end{array}
\end{equation}

To solve its dual problem in \eqref{eq11}, we should solve the problem in \eqref{eq10} with fixed $\{ {\lambda _i}\}$, then update the Lagrangian dual variables $\{ {\lambda _i}\}$  by solving the dual problem in \eqref{eq11}. Iterate the above two steps until convergence.

\subsubsection{Solving the dual function of \eqref{eq10}}

For given $\{ {\lambda _i}, \forall i\}$, solving the dual function is the same as solving the following optimization problem
\begin{equation}
\begin{array}{l}
\label{eq12}
\mathop {{\rm{max}}}\limits_{\{ {\bf{Q}}\succeq {\bf{0}}\} } \;\;{\rm{lo}}{{\rm{g}}_2}\left| {{\bf{I}} + {\bf{H}}{\bf{Q}}{{\bf{H}}^{H}}} \right| - {\rm{tr}}\left( {{\bf{B}}{\bf{Q}}} \right),
\end{array}
\end{equation}
where ${\bf{B}} = \sum\nolimits_{i = 1}^A {{\lambda _i}{{\bf{B}}_i}}$. We then have the following lemma.

\textbf{Lemma 1: } For the problem in \eqref{eq12} to have a bounded objective value, matrix ${\bf{B}}$ should be positive definite.

\textbf{Proof:} Obviously, ${\bf{B}}\succeq {\bf{0}}$. The remaining task is to show that ${\bf{B}} \ne {\bf{0}}$. If ${\bf{B}} = {\bf{0}}$, the objective value of the problem in  \eqref{eq12} will grow without bound. Therefore, ${\bf{B}} \succ {\bf{0}}$ holds. \hfill $\Box$

According to Lemma 1, ${\bf{B}}$ can be written as ${\bf{B}} = {{\bf{B}}^{1/2}}{{\bf{B}}^{1/2}}$. Let ${\bf{\tilde Q}} = {{\bf{B}}^{1/2}}{\bf{Q}}{{\bf{B}}^{1/2}}$ , we can get ${\bf{Q}} = {{\bf{B}}^{ - 1/2}}{\bf{\tilde Q}}{{\bf{B}}^{ - 1/2}}$. Inserting it into \eqref{eq12} yields
\begin{equation}
\begin{array}{l}
\label{eq13}
\mathop {{\rm{max}}}\limits_{{\bf{\tilde Q}}\succeq{\bf{0}}} \;{\rm{lo}}{{\rm{g}}_2}\left| {{\bf{I}} + {\bf{\tilde Q}}{{\bf{B}}^{ - \frac{1}{2}}}{{\bf{H}}^H}{\bf{H}}{{\bf{B}}^{ - \frac{1}{2}}}} \right| - {\rm{tr}}\left( {{\bf{\tilde Q}}} \right).
\end{array}
\end{equation}
To solve this problem, we first write the eigenvalue decomposition (EVD)  of ${{\bf{B}}^{ - 1/2}}{{\bf{H}}^H}{\bf{H}}{{\bf{B}}^{ - 1/2}}$ as
\begin{equation}\label{eigvector}
\begin{array}{l}
{{\bf{B}}^{ - \frac{1}{2}}}{{\bf{H}}^H}{\bf{H}}{{\bf{B}}^{ - \frac{1}{2}}} = {\bf{UD}}{{\bf{U}}^{\rm{H}}},
\end{array}
\end{equation}
where ${\bf{U}}$ is a semi-unitary matrix of the eigenvectors with
\begin{equation}
\begin{array}{l}
r \buildrel \Delta \over = {\rm{rank}}\left( {{{\bf{B}}^{ - \frac{1}{2}}}{{\bf{H}}^H}{\bf{H}}{{\bf{B}}^{ - \frac{1}{2}}}} \right),
\end{array}
\end{equation}
and ${\bf{D}}$  is a diagonal matrix with ${\bf{D}} = {\rm{diag}}\{ {d_1}, \cdots ,{d_r}\} $ being the eigenvalues.

Then, the solution to the problem in \eqref{eq13} is given by
\begin{equation}
\begin{array}{l}
{\bf{\tilde Q}} = {\bf{U}}\Lambda {{\bf{U}}^{\rm{H}}},
\end{array}
\end{equation}
where $\Lambda  = {\rm{diag}}\{ {q_1}, \cdots ,{q_r}\} $ represents the power allocation on all subchannels with
\begin{equation}
\begin{array}{l}
{q_m} = {\left[ {\frac{1}{{\ln 2}} - \frac{1}{{{d_m}}}} \right]^ + },m = 1, \cdots ,r,
\end{array}
\end{equation}
where ${[x]^ + } = \max \{ 0,x\} $.

Hence, the optimal ${\bf{Q}}$  is given by
\begin{equation}\label{optimalprecoding}
\begin{array}{l}
{\bf{Q}} = {{\bf{B}}^{ - \frac{1}{2}}}{\bf{U}}\Lambda {{\bf{U}}^{\rm{H}}}{{\bf{B}}^{ - \frac{1}{2}}}.
\end{array}
\end{equation}

\subsubsection{Solving the dual problem  \eqref{eq11}}

To solve the dual problem  \eqref{eq11}, we utilize the subgradient method, which is a simple method to deal with the nondifferentiable objective function \cite{p4-palomar2006tutorial}. The subgradient is required by the subgradient method at each iteration. According to \cite{p4-palomar2006tutorial}, the subgradient of $g( \cdot )$ at ${\lambda ^{(k)}} = [{\lambda _1}, \cdots ,{\lambda _I}]$ in the $k{\rm{th}}$ iteration is given by \cite{p4-palomar2006tutorial}
\begin{equation}\label{subgrad}
\begin{array}{l}
s_i^{(k)} = {P_i} - {\rm{tr}}\left( {{{\bf{B}}_i}{\bf{Q}}} \right),i = 1, \cdots ,I.
\end{array}
\end{equation}
Then the Lagrangian dual variables can be updated as
\begin{equation}\label{dualvariabel}
\begin{array}{l}
\lambda _i^{(k + 1)} = \left[ {\lambda _i^{(k)} - {u^{(k)}}s_i^{(k)}} \right],\forall i.
\end{array}
\end{equation}
where ${u^{(k)}}$ is the step in the $k{\rm{th}}$ iteration. The subgradient method is guaranteed to converge if ${u^{(k)}}$ satisfies ${\lim _{k \to \infty }}{u_k} = 0$ and $\sum\nolimits_{k = 1}^\infty  {{u_k}}  = \infty $  \cite{p5-boyd2003subgradient}.

In summary, the algorithm to solve Problem $(\rm{\textbf{P1}})$ is given in Algorithm 2.

\begin{rem}
For the special case of $N=1$, the channel matrix ${\bf{H}}$ reduces to a vector ${\bf{h}}$ and the semi-unitary matrix ${\bf{U}}$ in (\ref{eigvector})  reduces to a vector, given by
\begin{equation}\label{you}
{\bf{u}} = \frac{{{{\bf{B}}^{ - 1/2}}{\bf{h}}}}{{\left\| {{{\bf{B}}^{ - 1/2}}{\bf{h}}} \right\|}}.
\end{equation}
Also, diagonal matrix ${\bf{D}}$ reduces to a scalar $q$ and only one subchannel is used to transmit information, that is power allocation matrix $\Lambda$ reduces to a scalar $q$.  Hence, according to (\ref{optimalprecoding}), the optimal transmit convariance matrix is given by
\begin{equation}\label{precoding}
 {\bf{Q}} = \frac{q}{{\left\| {{{\bf{B}}^{ - 1/2}}{{\bf{h}}^H}} \right\|^2}}{{\bf{B}}^{ - 1}}{{\bf{h}}^H}{\bf{h}}{{\bf{B}}^{ - 1}}.
\end{equation}
Since ${\bf{B}}$ is a diagonal matrix, the optimal transmission strategy should be the maximum ratio transmission, which is consistent with the analysis in \cite{Sang-2013}. Here, we have extended the result in \cite{Sang-2013} to the more general case with multiple antennas at the receiver.
\end{rem}

\begin{algorithm}
\caption{Solving Problem $(\rm{\textbf{P1}})$}

\textbf{Initialize:} Iteration number $k = 0$, ${\lambda ^{(0)}} = [{\lambda _1}, \cdots ,{\lambda _A}]$, such that ${\bf{B}} \succ {\bf{0}}$.

\textbf{Repeat}

  \quad 1. Compute ${\bf{Q}}(\lambda ) = {{\bf{B}}^{ - \frac{1}{2}}}{\bf{U}}\Lambda {{\bf{U}}^{\rm{H}}}{{\bf{B}}^{ - \frac{1}{2}}}$ with fixed ${\lambda ^{(k)}}$ using (\ref{optimalprecoding});

  \quad 2. Compute the subgradient $s_i^{(k)},\forall i$, using (\ref{subgrad});

  \quad 3. Update $\lambda _i^{(k + 1)}$ using (\ref{dualvariabel}), increase $k$ by $1$;

\textbf{Until} convergence

\end{algorithm}

\subsection{Algorithm to solve Problem $(\rm{\textbf{P2}})$}

Since the numerator and denominator in the objective function of Problem $(\rm{\textbf{P2}})$ are concave and affine in ${\bf{Q}}$ respectively, the objective function is a pseudo-concave function \cite{p7-zappone2014energy}. In addition, the available power region is a convex set. Hence, Problem $(\rm{\textbf{P2}})$ can be efficiently solved by using the following lemma, the proof of which can be found in \cite{p8-dinkelbach1967nonlinear}.

\textbf{Lemma 2}: Define function $G(\eta )$ as
\begin{equation}\label{fractional}
\begin{array}{l}
G(\eta ) \buildrel \Delta \over = \mathop {\rm{max}}\limits_{{{\bf{Q}}} \in {\cal W}} \;\;{\log _2}\left| {{\bf{I}} + {\bf{H}}{{\bf{Q}}}{{\bf{H}}^{H}}} \right| - \eta \left( {{\rm{tr}}\left( {{{\bf{Q}}}} \right){\rm{ + }}{P_C}} \right),
\end{array}
\end{equation}
For a fixed $\eta $, the solution to the problem in (\ref{fractional}) is denoted as ${\bf{Q}}^*(\eta )$. Then solving Problem $(\rm{\textbf{P2}})$ is equivalent to finding the root of the equation $\eta^*$, which satisfies

\begin{equation}
\begin{array}{l}
\label{eq27}
G({\eta ^*})\! =\! {{\textmd{log}}_2}\left| {{\bf{I}}\! +\! {{\bf{H}}}{\bf{Q}}^*({\eta ^*}){\bf{H}}^H} \right|\! - \!{\eta ^*}\!\left( {{\rm{tr(}}{{\bf{Q}}}^*({\eta ^*}){\rm{) \!+ \! }}{P_C}} \!\right)\! =\! 0.
\end{array}
\end{equation}

Lemma 2 gives us insights to solve Problem $(\rm{\textbf{P2}})$. We should firstly solve the problem in (\ref{fractional}) for a fixed $\eta $, and then utilize the Dinkelbach method \cite{p8-dinkelbach1967nonlinear} to update $\eta $. For a fixed $\eta $,  the solution to the problem in \eqref{fractional} can be obtained by the same method as the one used to solve Problem $(\rm{\textbf{P1}})$, except that ${\bf{B}}$ is replaced by ${\bf{\tilde B}} = \eta {\bf{I}} + \sum\nolimits_{i = 1}^{A} {{\lambda _i}{{\bf{B}}_i}} $.

After solving the problem in \eqref{fractional}, we utilize the Dinkelbach method \cite{p8-dinkelbach1967nonlinear} to update $\eta $ as follows
\begin{equation}
\begin{array}{l}
\label{eq28}
{\eta ^{(n + 1)}} = \frac{{{\textmd{log}_2}\left| {{\bf{I}} + {\bf{H}\bf{Q}}^*({\eta ^{(n)}}){{\bf{H}}^{\bf{H}}}} \right|}}{{{\rm{tr}}\left( {{\bf{Q}}^*({\eta ^{(n)}})} \right){\rm{ + }}{P_C}}},
\end{array}
\end{equation}
where $n$  is the iteration index.

Based on the above analysis, we provide Algorithm 3 to solve Problem $(\rm{\textbf{P2}})$.

\begin{algorithm}
\caption{The Dinkelbach method to solve Problem $(\rm{\textbf{P2}})$}

\textbf{Initialize:} ${\eta ^{(0)}}$ satisfying $G({\eta ^{(0)}}) \ge 0$,  iteration number $n = 0$;

\textbf{Repeat}

  \quad 1. Compute the optimal solution ${\bf{Q}}^*({\eta ^{(n)}})$ with fixed ${\eta ^{(n)}}$ by using Algorithm 2 except that ${\bf{B}}$ is replaced by ${\bf{\tilde B}} = \eta {\bf{I}} + \sum\nolimits_{i = 1}^{A} {{\lambda _i}{{\bf{B}}_i}} $.

  \quad 2. Update ${\eta ^{(n + 1)}}$ according to \eqref{eq28}, and set $n \leftarrow n + 1$;

\textbf{Until} convergence
\end{algorithm}

 If the optimal solution to Problem $(\rm{\textbf{P2}})$ does not satisfy the rate requirement, save the obtained ${\eta^*}$ from Algorithm 3, which will be used to solve Problem $(\rm{\textbf{P3}})$ as shown in the next subsection.

\subsection{Algorithm to solve Problem $(\rm{\textbf{P3}})$}

Problem $(\rm{\textbf{P3}})$ is non-convex due to the rate \emph{equality} constraint. However, it is straightforward to see that the solution is equal to the following convex problem
\begin{equation}
\begin{array}{*{20}{l}}
{\;\mathop {{\rm{min}}}\limits_{{\bf{Q}} \in W} \;\;{\rm{tr}}\left( {\bf{Q}} \right)}\\
{\;\;\;{\rm{s.t.}}\;\;\;{\rm{lo}}{{\rm{g}}_{\rm{2}}}\left| {{\bf{I}} + {\bf{HQ}}{{\bf{H}}^{\bf{H}}}} \right| \ge {{\tilde R}_{\min }},\;}\\
{\;\;\;\;\;\;\;\;\;\;\;\;}
\end{array}
\end{equation}
Thus, it can be solved by the Lagrangian dual method.

The dual function is given by
\begin{equation}
\begin{array}{l}
\label{eq31}
g(\lambda ) = \mathop {{\rm{min}}}\limits_{ {{\bf{Q}}} \in {\cal W} } \;{\rm{tr}}\left( {{{\bf{Q}}}} \right) + \lambda \left( {{\tilde R_{\min }} - {\rm{log_2}}\left| {{\bf{I}} + {\bf{H}}{{\bf{Q}}}{{\bf{H}}^{\bf{H}}}} \right|} \right).
\end{array}
\end{equation}
Then the dual problem is
\begin{equation}
\begin{array}{l}
\mathop {\max }\limits_{\lambda  \ge 0} g(\lambda ).
\end{array}
\end{equation}
Now, we solve the problem in \eqref{eq31} for a given $\lambda$. If $\lambda  = 0$, the optimal ${{\bf{Q}}}$ should be the zero matrix. In this case, the rate becomes zero, which cannot be the optimal dual variable. Hence, $\lambda $ should be positive.  Denote $\mu  = 1/\lambda  > 0$, the problem in \eqref{eq31} can be equivalently expressed as
\begin{equation}
\begin{array}{l}
\label{eq33}
\mathop {{\rm{max}}}\limits_{ {\bf{Q}} \in {\cal W} } \;\;{\rm{log_2}}\left| {{\bf{I}} + {\bf{H}}{{\bf{Q}}}{{\bf{H}}^{H}}} \right| - \mu {\rm{tr}}\left( {{{\bf{Q}}}} \right).
\end{array}
\end{equation}
 Note that for a fixed $\mu$, the problem in \eqref{eq33} is the same as the problem in \eqref{fractional}, and thus can be solved by the same method as the one used to solve Problem $(\rm{\textbf{P1}})$. Denote the optimal solution of the problem in (\ref{eq33}) as ${\bf{Q}}^*(\mu )$. According to the complementary slackness condition \cite{boyd2004convex}, we should find the optimal ${\mu ^*}$ such that ${\rm{log_2}}\left| {{\bf{I}} + {{\bf{H}}{\bf{Q}}}^*({\mu ^*}){{\bf{H}}^{H}}} \right| = {\tilde R_{\min }}$.  
In the following lemma, we show that the optimal ${\mu ^*}$ should be smaller than the ${\eta ^*}$ obtained in Algorithm 3.

\textbf{Lemma 3: } The optimal ${\mu ^*}$ that leads to ${\rm{log_2}}\left| {{\bf{I}} + {{\bf{H}}{\bf{Q}}}^*({\mu ^*}){{\bf{H}}^{H}}} \right| = {\tilde R_{\min }}$ is smaller than the ${\eta ^*}$  obtained in Algorithm 3.

\textbf{proof:} See Appendix A. \hfill $\Box$

Based on Lemma 3, we can use the bisection search method to find the optimal ${\mu ^*}$ with the upper bound given by ${\eta ^*}$.

\subsection{Complexity Analysis}

In this subsection, we provide the complexity analysis for the EE optimization problem (\ref{EEwithfixedset}) for a fixed $\mathbb{S}$. Since this problem is divided into three subproblems, we should provide the complexity for each of these subproblems.  We assume that ${M_A}\geq N$. For Algorithm 2  to solve Problem $(\rm{\textbf{P1}})$, the main complexity lies in step 1, where the EVD operation in (\ref{eigvector}) is performed. According to \cite{boyd2004convex}, this operation incurs the complexity of $2M_A^2N+2N^3/3$ flops. Hence, the total computational complexity of Algorithm 2 involves $T_{\rm{alg2}}(2M_A^2N+2N^3/3)$ flops, where $T_{\rm{alg2}}$ is the average number of iterations required to converge. The simulation results show that Algorithm 2 converges fast with the proper choice of the step size. For Problem $(\rm{\textbf{P2}})$, the Dinkelbach method is used, in which the problem in (\ref{fractional}) is solved in each iteration that has a similar complexity as that of the Problem $(\rm{\textbf{P1}})$. Hence, the total complexity of Algorithm 3 involves $T_{\rm{alg3}}T_{\rm{alg2}}(2M_A^2N+2N^3/3)$ flops, where $T_{\rm{alg3}}$ is the number of iterations required for Algorithm 3 to converge. It is seen from the simulation results in Section \ref{simulation} that only a few number of iterations are enough for Algorithm 3 to converge. As for Problem $(\rm{\textbf{P3}})$, the bisection search method is employed which involves solving the problem in (\ref{eq33}) in each iteration that has a similar complexity as that for $(\rm{\textbf{P1}})$. Thus, the total complexity to solve $(\rm{\textbf{P3}})$ involves $T_{\rm{bis}}T_{\rm{alg2}}(2M_A^2N+2N^3/3)$ flops, where $T_{\rm{bis}}$ denotes the average number of iterations for the bisection search method to converge. Simulation results show that  $T_{\rm{bis}}$ is very small. Hence, the total complexity to solve the problem in (\ref{EEwithfixedset}) involves at most $(T_{\rm{alg3}}+T_{\rm{bis}}+1)T_{\rm{alg2}}(2{M_A}^2N+2N^3/3)$ flops.

 On the other hand, the complexity of the interior-point method \cite{boyd2004convex} is ${(m + 1)^{1/2}}n({n^2} +\sum\nolimits_{i = 1}^m {k_i^3} + n\sum\nolimits_{i = 1}^m {k_i^2} )$ flops \cite{Ben2001}, where $n$ is the number of unknown variables, $m$ is the number of constraints, and $k_i$ is the constant length in the $i{\rm{th}}$ constraint\cite{Ben2001}. For the problem in (\ref{EEwithfixedset}),  $n=M_{\rm{A}}^2$, $m=A+1$ and $k_i=1,\forall i$. In general, the number of receive antennas at the user is much smaller than that of the active antennas, i.e., $N\ll M_{\rm{A}}$.
 Hence, our algorithm yields a much lower computational complexity than the interior-point method.

\section{Active RAU Selection}\label{portselection}

In this section, we study the active RAU selection scheme. Obviously, the optimal RAU selection can be obtained by exhaustive search. Specifically, we compute the EE for all possible RAU selection  sets $\mathbb{S} \subseteq \left\{ {1,2, \cdots ,I} \right\}$ based on the algorithms developed in Section \ref{EEfixedsec}. Then, we select the best RAU set which has the best EE performance\footnote{ If problem (\ref{EEwithfixedset}) is infeasible, we set the corresponding EE to be zero. If there is no feasible set, we activate all transmit antennas to achieve the best sum rate performance. }. However, the complexity of exhaustive search is too high to implement in practice. Hence, a more efficient RAU selection scheme is required.

In \cite{Jie2013}, the authors proposed a norm-based antenna selection scheme for the centralized antenna system. It can be extended to the RAU selection here. In particular, we need to calculate the norms of the channel matrices from all RAUs to the user and sort them in decreasing order. Then, the RAU is added in the active RAU set one after another based on the sorted order.

However, in the DAS, the distances from the user to the RAUs are different and the channel norm is mainly determined by the distance. Based on these facts, we only need to obtain the distance information\footnote{According to \cite{Heejin2012}, the distance information can be easily calculated through the GPS or other similar methods.} and sort them in increasing order instead of calculating the channel norms. This method can reduce the computational complexity. It will be shown in Section \ref{simulation} that the EE performance of these two selection schemes are almost the same.  Denote the distance from the $i{\rm{th}}$ RAU to the user as $d_i$, the distance-based RAU selection scheme is given in Algorithm 4.

\begin{algorithm}
\caption{Distance-Based RAU Selection Scheme}
\textbf{Initialize:} Set ${\rm{EE}}_{\rm{opt}}=0$, feasibility mark ${\rm{flag}}=0$, the optimal RAU set  $\mathbb{S}_{\rm{opt}}=\emptyset$, the number of active RAUs $A=1$. Sort the distances $d_i$s in increasing order, i.e, ${d_{\pi (1)}} \le  \cdots  \le {d_{\pi (I)}}$, where $\pi (i)$ represents the index of the user that has the $i$th smallest distance.
\begin{enumerate}
  \item  Set $\mathbb{S}_a=\{\pi (1),\cdots,\pi (A) \}$ and the channel matrix from the selected RAUs as  ${{\bf{H}}_ {\mathbb{S}_a} } = \left[ {{{\bf{H}}_{{\pi (1)}}},{{\bf{H}}_{{\pi (2)}}}, \cdots ,{{\bf{H}}_{{\pi (A)}}}} \right]$.
  \item Solve the EE optimization problem in (\ref{EEwithfixedset}) by using Algorithm 1 based on the active RAU set $\mathbb{S}_a$. If it is infeasible, set ${\rm{EE}}_{\mathbb{S}_a}=0$. Otherwise, set ${\rm{flag}}=1$ and compute the achieved EE ${\rm{EE}}_{\mathbb{S}_a}$.
  \item  If ${\rm{EE}}_{\rm{opt}}\leq{\rm{EE}}_{\mathbb{S}_a}$, set ${\rm{EE}}_{\rm{opt}}={\rm{EE}}_{\mathbb{S}_a}$ and $\mathbb{S}_{\rm{opt}}={\mathbb{S}_a}$; Otherwise, terminate.
  \item  If $A<I$, set $A=A+1$ and go to Step 1); Otherwise, go to Step 5),
  \item  If ${\rm{flag}}=0$, set $\mathbb{S}_{\rm{opt}}=\{1,\cdots,I\}$, solve the rate maximization problem with Algorithm 2 based on $\mathbb{S}_{\rm{opt}}$, compute the corresponding EE ${\rm{EE}}_{\rm{opt}}$, and terminate; Otherwise, terminate.
\end{enumerate}
\end{algorithm}

Compared with the antenna selection algorithm in \cite{Jie2013}, our proposed algorithm does not need to solve the EE optimization problem $I$ times. It will be terminated as long as the current achieved EE is smaller than the previous highest EE as seen in step 3) of Algorithm 4. This can significantly reduce the search complexity especially when the number of RAUs is large. This is also reasonable,  because if adding one more RAUs in the selection RAU set decreases the EE, the remaining RAUs will also degrade the EE since their distances are much larger and the corresponding channel gains are much smaller. The simulation results will demonstrate this statement.


We compare the complexity of our proposed algorithm with the exhaustive search. In the exhaustive search, the size of the set of RAU candidates $\mathbb{S}$ is given as ${2^I} - 1$, and thus the problem in (\ref{EEwithfixedset}) should be solved ${2^I} - 1$ times, which increases exponentially with the RAU number $I$. On the other hand, for our proposed RAU selection algorithm, i.e., Algorithm 4, we solve the problem in (\ref{EEwithfixedset}) at most $I$ times. 
For example, for $I=20$, 1048575 RAU selection sets are required for the exhaustive search, while at most 20 candidates are required in our proposed selection scheme, which account for only 0.19\textperthousand \  of the complexity of the exhaustive search.

\section{Simulation Result}\label{simulation}

In this subsection, we evaluate the performance of the proposed RAU selection and transmit covariance optimization scheme, i.e., Algorithm 4, through simulations. We assume the user is uniformly distributed in the circular cell centered at $(0,0)$ with radius $R$. According to \cite{Yunzhi-2009} and \cite{Xinzheng-2009}, if the number of RAUs $I$ is smaller than 6, the location of the $j$-th RAU is $\left( {r\cos ({{2\pi (j - 1)} \mathord{\left/
 {\vphantom {{2\pi (j - 1)} I}} \right.
 \kern-\nulldelimiterspace} I}),r\sin ({{2\pi (j - 1)} \mathord{\left/
 {\vphantom {{2\pi (j - 1)} I}} \right.
 \kern-\nulldelimiterspace} I})} \right)$ for $j=1,\cdots, I$, where $r = {{2R\sin ({\pi  \mathord{\left/
 {\vphantom {\pi  I}} \right.
 \kern-\nulldelimiterspace} I})} \mathord{\left/
 {\vphantom {{2R\sin ({\pi  \mathord{\left/
 {\vphantom {\pi  I}} \right.
 \kern-\nulldelimiterspace} I})} {\left( {{{3\pi } \mathord{\left/
 {\vphantom {{3\pi } I}} \right.
 \kern-\nulldelimiterspace} I}} \right)}}} \right.
 \kern-\nulldelimiterspace} {\left( {{{3\pi } \mathord{\left/
 {\vphantom {{3\pi } I}} \right.
 \kern-\nulldelimiterspace} I}} \right)}}$ . Otherwise, the first RAU is located at the cell center $(0,0)$, and the other $I-1$ RAUs are located at $\left( {r\cos ({{2\pi (j - 2)} \mathord{\left/
 {\vphantom {{2\pi (j - 2)} {(I - 1)}}} \right.
 \kern-\nulldelimiterspace} {(I - 1)}}),r\sin ({{2\pi (j - 2)} \mathord{\left/
 {\vphantom {{2\pi (j - 2)} {(I - 1)}}} \right.
 \kern-\nulldelimiterspace} {(I - 1)}})} \right)$ with $j=2,\cdots,I$. The channel modeled considers large-scale pathloss \cite{Assumptions-2009},  shadow fading and independent Rayleigh fading. More specifically, the shadow fading follows the log-normal distribution with the
standard deviation $\delta_{sh}=8 {\rm{dB}}$. For simplicity, we assume each RAU has the same number of antennas $M_j=M,j=1,\cdots,I$ and the same power constraints $P_j=P_{\rm{max}},j=1,\cdots,I$. Unless otherwise  specified, the other main system parameters are given in Table \ref{tab1}. The following results are obtained by averaging over 500 independent channel generations.

 \begin{table}[h]
\renewcommand{\arraystretch}{1.1}
\caption{\label{parameter}Main simulation parameters}
\label{tab1}
\centering
\begin{tabular}{c|c}
\hline\hline
\textbf{Parameters}  & \textbf{Value} \\
\hline
Channel bandwidth $W$  & 20 MHz \\
\hline
Cell radius $R$   &  1000 $m$\\
\hline
Number of transmit antennas  $M$  & 4 \\
\hline
Number of receiver antennas $N$   &  4 \\
\hline
 Noise power spectral density  & -174 dBm/Hz \\
\hline
RF chain power consumption $p_c$ & 1 W\\
\hline
Static power consumption $p_0$  & 1 W\\
\hline
Tolerance  $\varepsilon$ &  $10^{(-5)}$  \\
\hline
Maximum transmit power  $P_{\rm{max}}$ &   10 W  \\
\hline
 Channel path loss model & $38.46 + 35{\log _{10}}(d)$ \cite{Assumptions-2009} \\
\hline\hline
\end{tabular}
\end{table}

\subsection{Convergence behavior of the proposed algorithms}

We first study the convergence behaviors of the proposed algorithms: Algorithm 2 to solve Problem $(\rm{\textbf{P1}})$, Algorithm 3 to solve Problem $(\rm{\textbf{P2}})$ and the bisection method to solve Problem $(\rm{\textbf{P3}})$. Since we only consider the convergence behaviours for these algorithms, we do not consider RAU selection in this subsection and all RAUs are turned on.

\begin{figure}[h]
\centering
\includegraphics[width=3.4in]{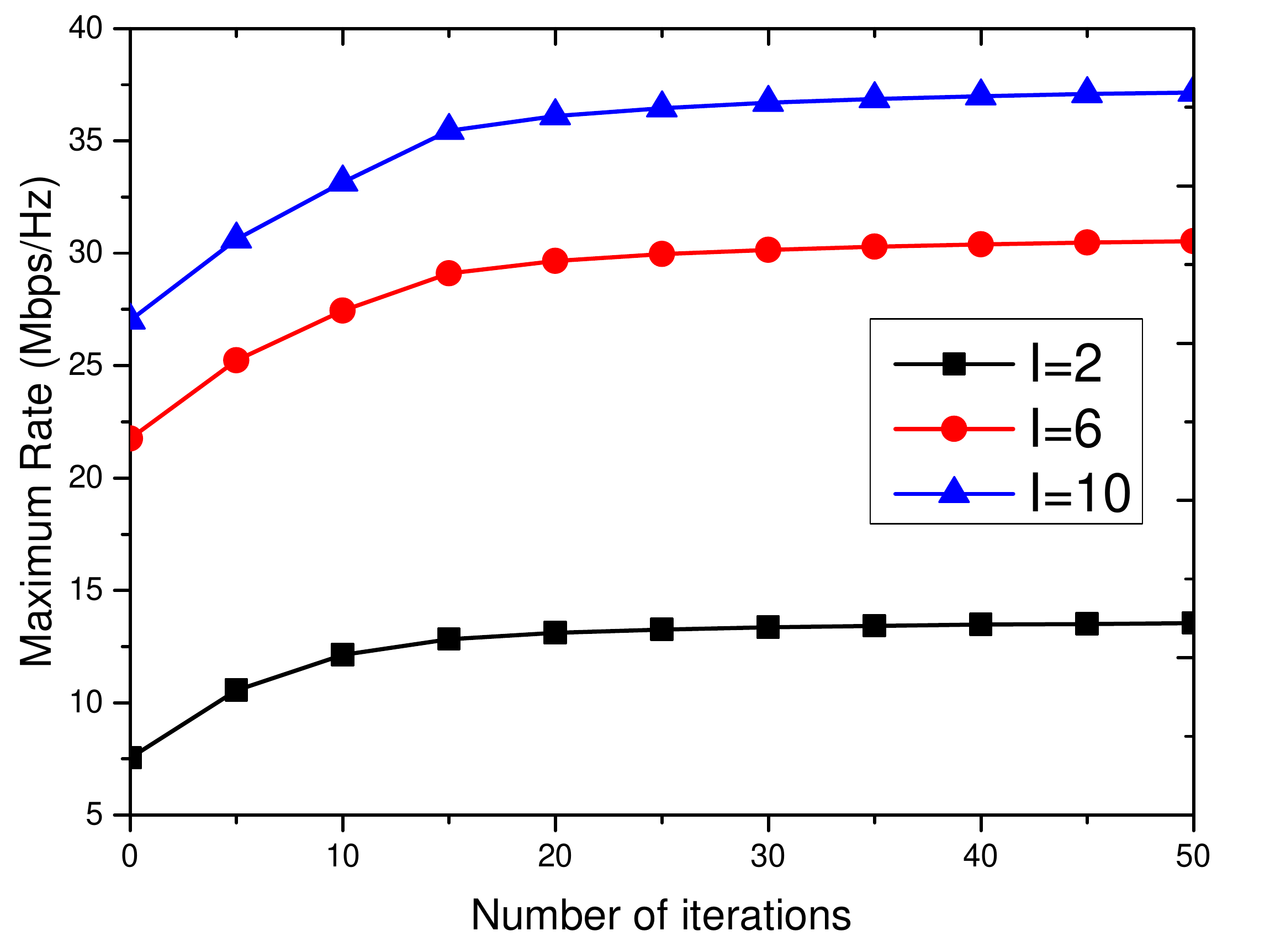}
\caption{Convergence behaviors of Algorithm 2 to solve Problem $(\rm{\textbf{P1}})$ under different number of RAUs.}
\label{convergenceP1}
\end{figure}

Fig. \ref{convergenceP1}  illustrates  the maximum rate achieved by Algorithm 2 versus the number of iterations under three different configurations: the DAS network with $I=2$,  $I=6$ and $I=10$. The step is set as ${u^{(k)}}=1/(30k)$, where $k$ is the iteration index. It can be seen from Fig. \ref{convergenceP1} that the maximum rate monotonically increases during the iterative procedure of Algorithm 2 and converges rapidly (within 50 iterations for all considered configurations).  On the other hand, from Fig. \ref{convergenceP1}, it can be seen that the maximum rate performance is significantly affected by the number of the antennas: a larger $I$ corresponds to  a higher rate.

\begin{figure}[h]
\centering
\includegraphics[width=3.4in]{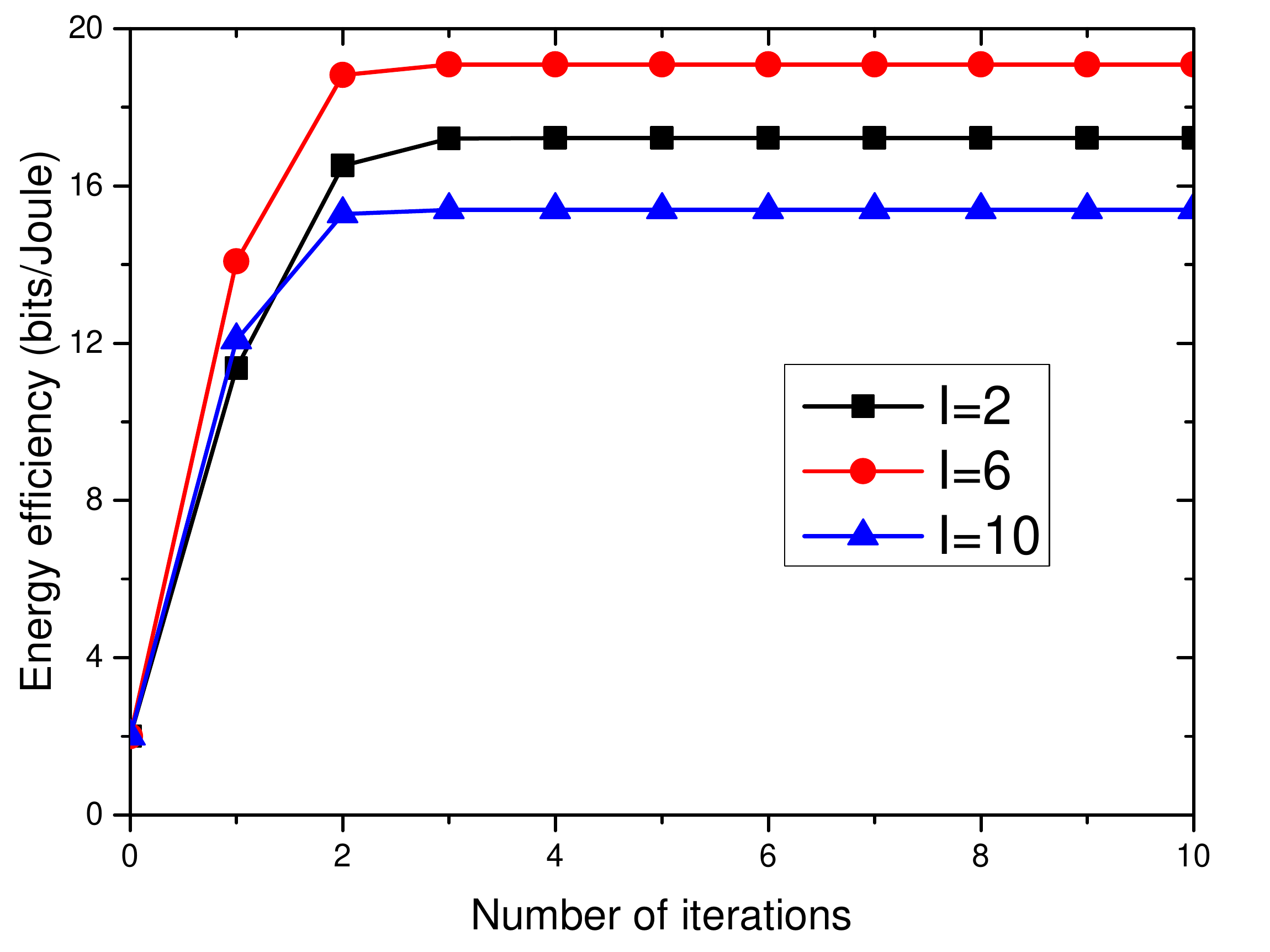}
\caption{Convergence behaviors of Algorithm 3 to solve Problem $(\rm{\textbf{P2}})$ under different number of RAUs.}
\label{convergenceP2}
\end{figure}

Fig. \ref{convergenceP2}  shows the convergence behaviour of Algorithm  3 for different number of antennas with $I=2$,  $I=6$ and $I=10$.
It can be seen from this figure that several iterations are enough for Algorithm 3 to converge. In addition,  a larger $I$ does not guarantee a higher EE due to more circuit power consumption. For example, the EE corresponding to $I=6$ is higher than the EE for the other two scenarios ($I=4$ and $I=10$). Hence, to achieve the highest EE, it is necessary to choose the appropriate set of active RAUs.

\begin{figure}[h]
\centering
\includegraphics[width=3.4in]{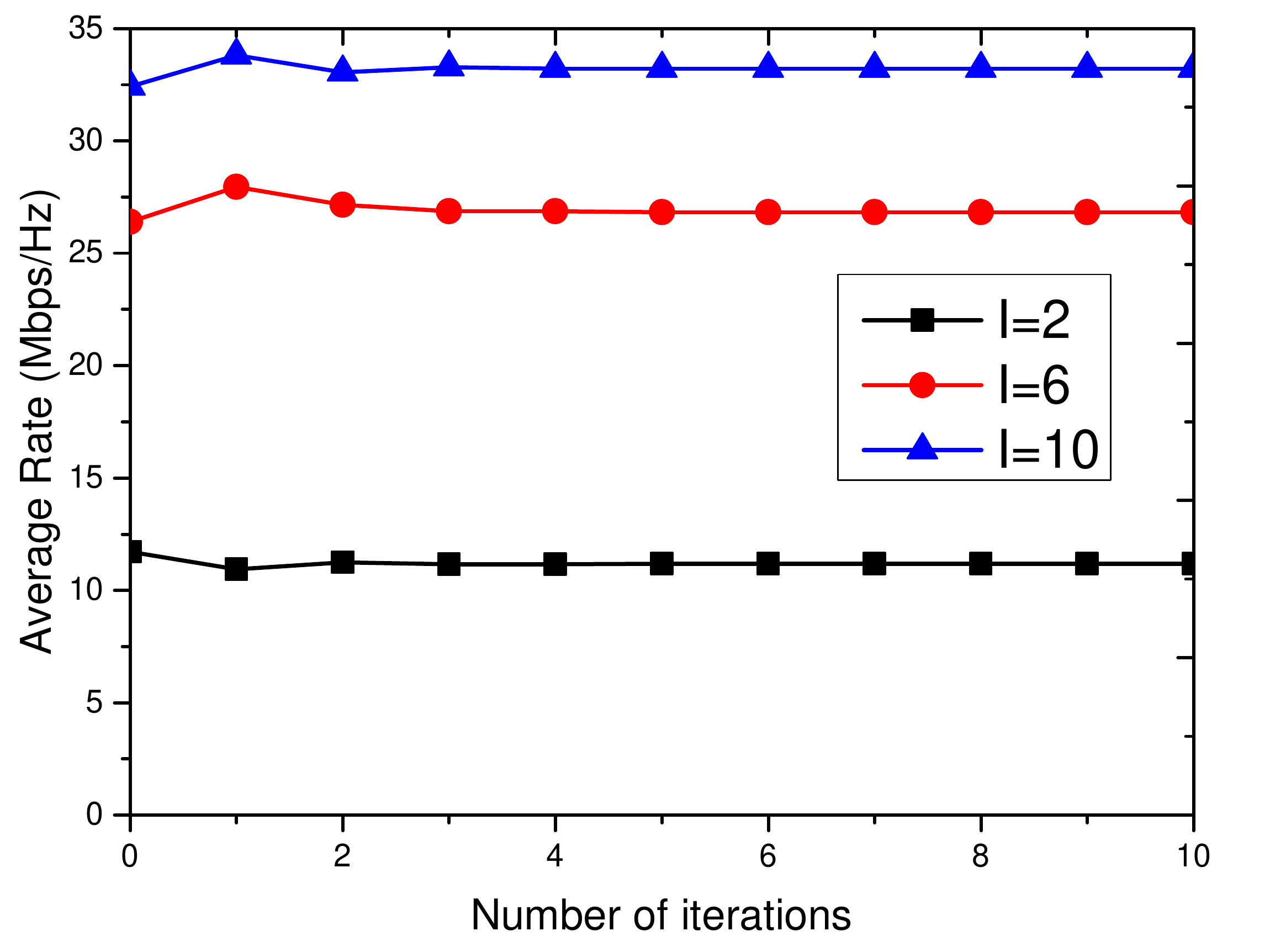}
\caption{Convergence behaviors of the bisection method to solve Problem $(\rm{\textbf{P3}})$ under different number of RAUs.}
\label{convergenceP3}
\end{figure}

The convergence behaviour of the bisection method to solve Problem $(\rm{\textbf{P3}})$ is given in Fig. \ref{convergenceP3}. Denote $R_{\rm{max}}$ and $R_{\rm{frac}}$ as the maximum rates achieved by solving Problem $(\rm{\textbf{P1}})$ and  Problem $(\rm{\textbf{P2}})$, respectively. In Problem $(\rm{\textbf{P3}})$, the rate constraint is set as $R_{\rm{min}}=(R_{\rm{max}}+R_{\rm{frac}})/2$.  It can be seen from Fig. \ref{convergenceP3} that the achievable  rate converges rapidly (within 10 iterations for all considered configurations).

\subsection{Performance comparison}

In this subsection, We compare our proposed RAU selection and transmit covariance optimization algorithm, i.e., Algorithm 4, with the following algorithms: exhaustive search which examines all possible RAU selection set (labeled as ``Exhaustive search''), norm-based RAU selection algorithm  (labeled as ``Prop norm-based alg.'') which is the same as Algorithm 4 except that we sort the channel norms in decreasing order, EE maximization without RAU selection and all available RAUs are active (labeled as ``EE without RAU sel.'') and SE maximization activating all available RAUs (labeled as ``SE maximization''). We also show the performance of the EE maximization proposed in \cite{Jie2013} for the centralized antenna system (CAS) where all antennas are placed at the center of the cell. For a fair comparison, we assume that the maximum transmit power for CAS is set to be $IP_{\rm{max}}$ and the base station is equipped with $IM$ antennas. Moreover, the total circuit power consumption of CAS is defined as $Ip_c+IMp_0$.

\begin{figure}[h]
\centering
\includegraphics[width=3.4in]{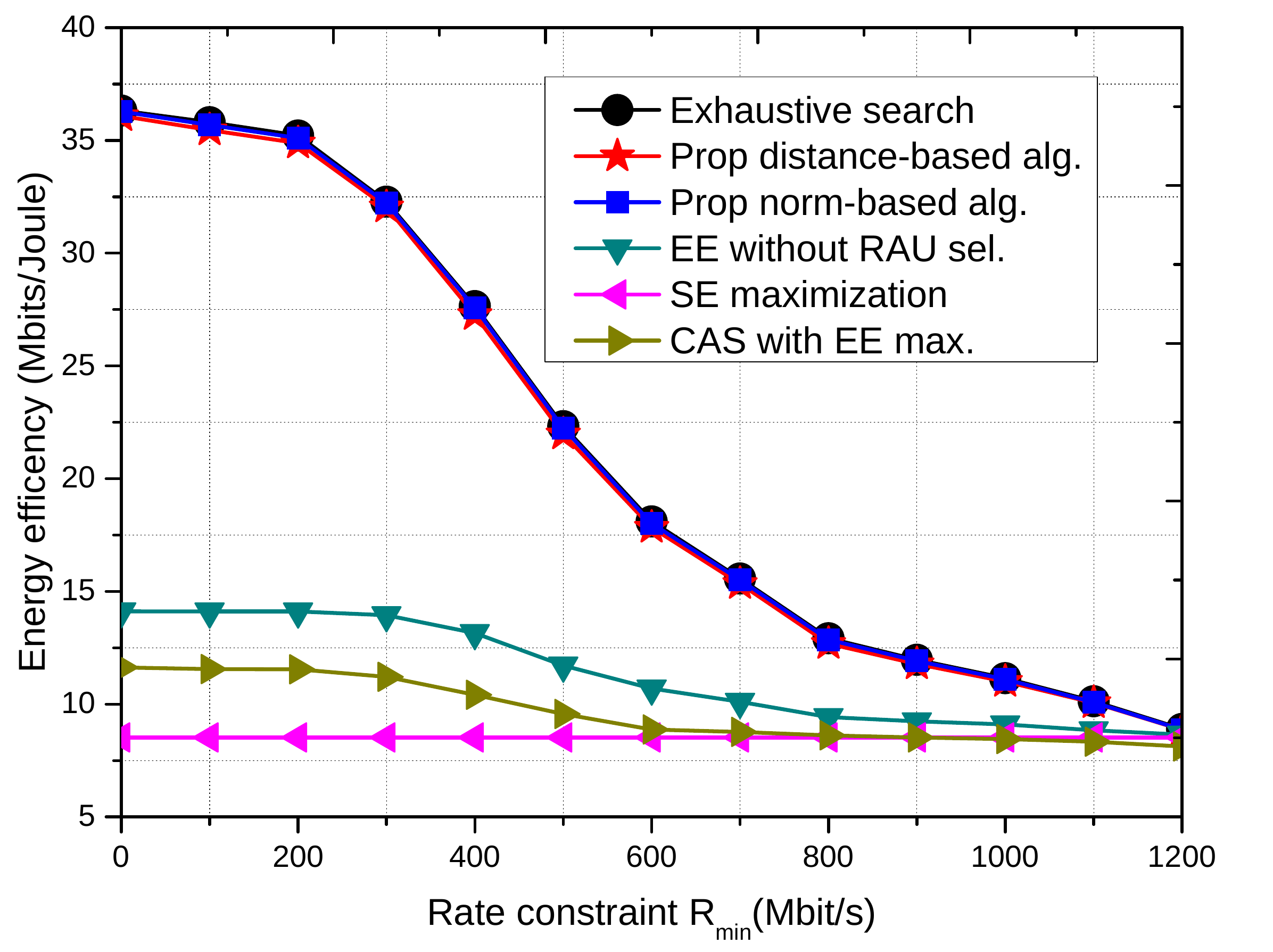}
\caption{EE under different rate constraints, where $I=4$.}
\label{EErateconstraint}
\end{figure}

\begin{figure}[h]
\centering
\includegraphics[width=3.4in]{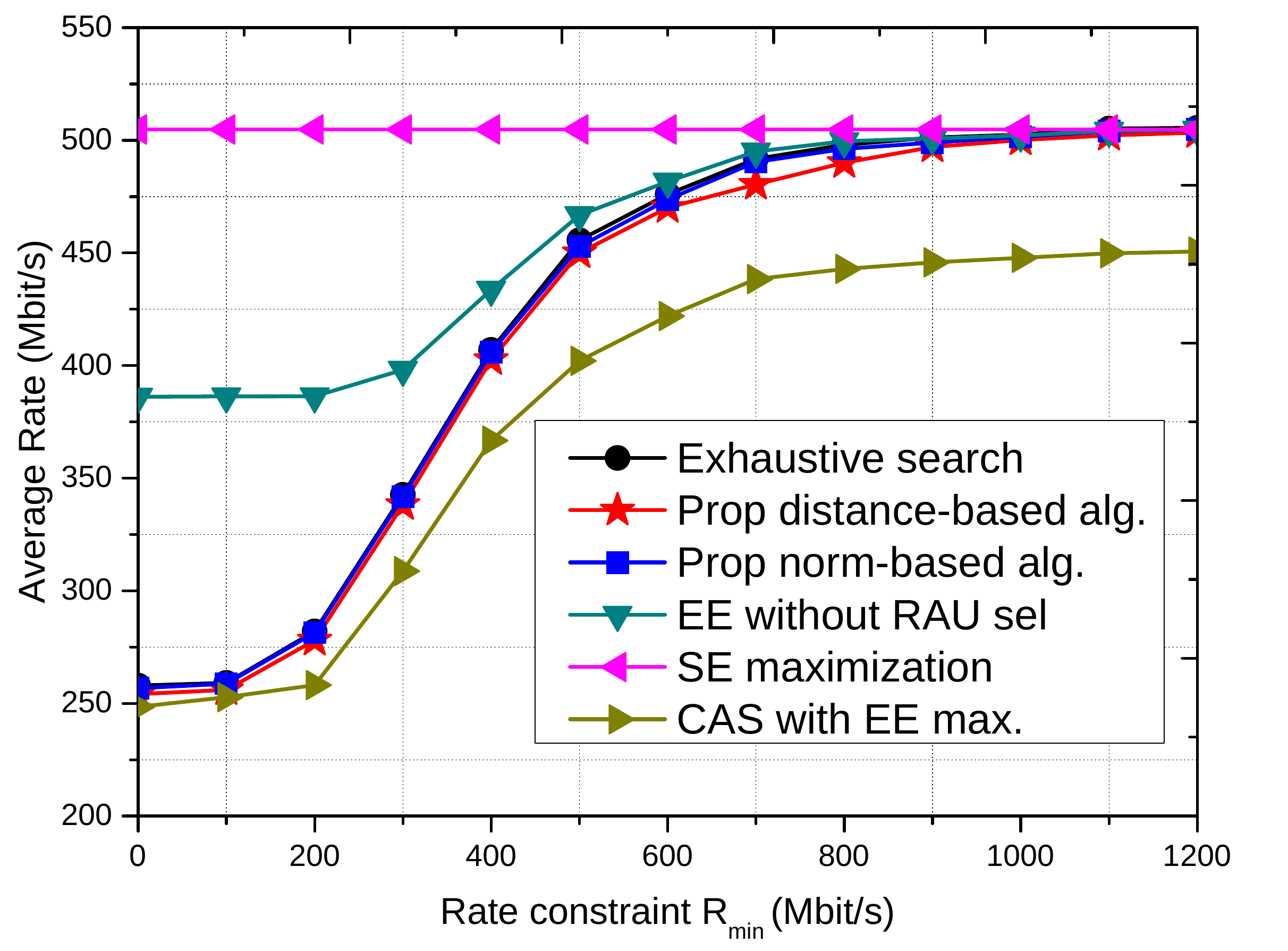}
\caption{Corresponding rate under different rate constraints, where $I=4$.}
\label{Raterateconstraint}
\end{figure}

\begin{figure}[h]
\centering
\includegraphics[width=3.4in]{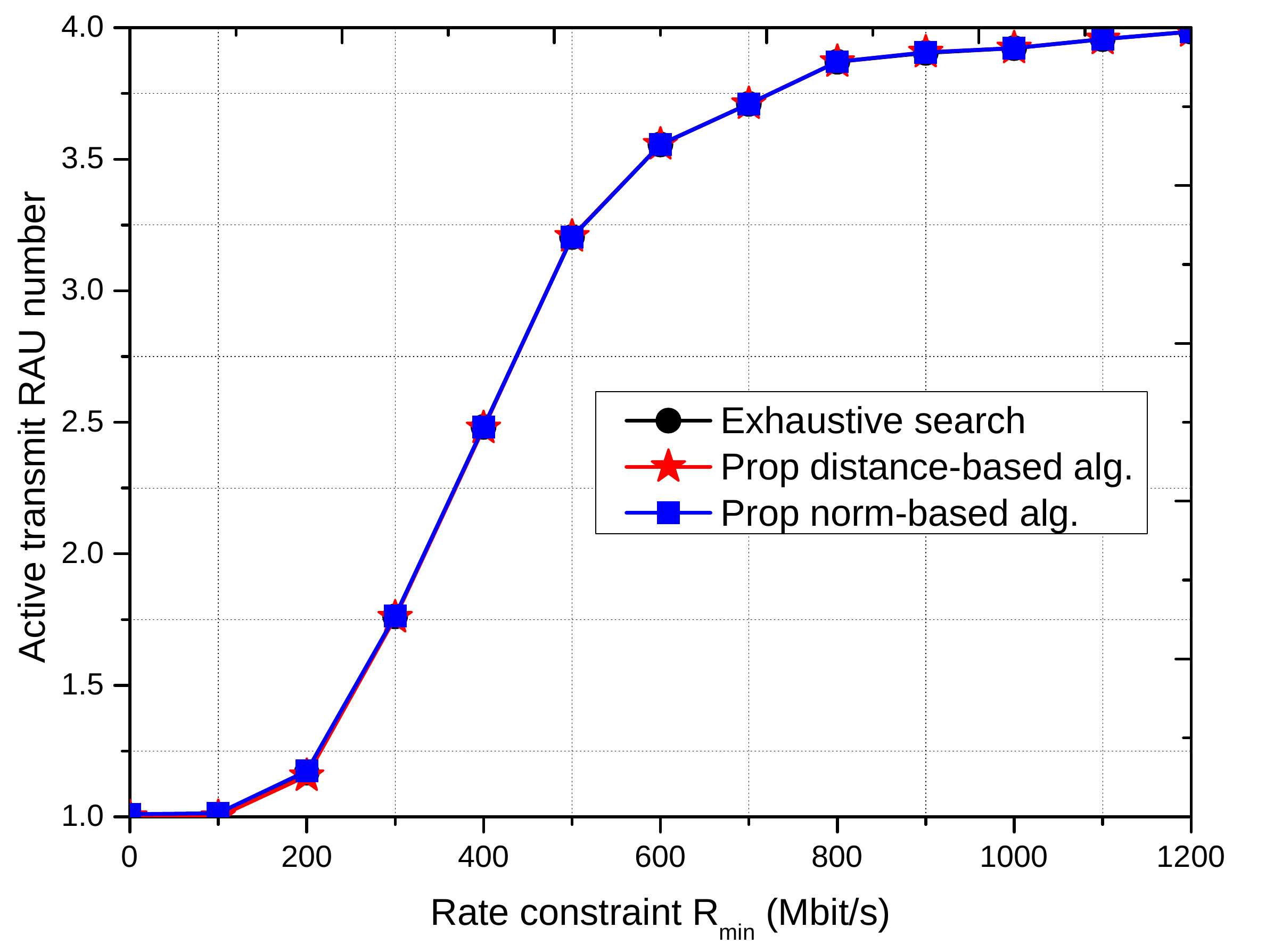}
\caption{Corresponding number of active RAUs under different rate constraints, where $I=4$.}
\label{Portrateconstraint}
\end{figure}

The effects of the rate constraint on the EE of the DAS are studied in Fig. \ref{EErateconstraint}, the corresponding rate and the number of active RAUs are shown in  Fig. \ref{Raterateconstraint} and Fig. \ref{Portrateconstraint}, respectively. The total number of RAUs is set as $I=4$. As expected,  the EE achieved by all algorithms decreases with the rate constraint. The reason is that,  more RAUs need to be active to ensure that the rate requirement can be satisfied, which incurs more circuit power consumption. Note that both the exhaustive search method and the channel-norm-based search method provide marginal gains over our proposed RAU selection method, and our proposed algorithm yields the least computational complexity. In addition, we observe that our proposed algorithm significantly outperforms the other schemes in terms of the EE, especially in the low rate constraint regime. The superiority of the proposed algorithm over the ``EE without RAU sel.'' comes from the RAU selection procedure. It means that turning off some RAUs can save significant amount of circuit power consumption when the rate requirement is satisfied, and thus improves the EE of the DAS.
It is interesting  to find that the EE performance of the DAS is much higher than that of the CAS. This observation is meaningful: to achieve the best EE performance, RAUs should be placed in a distributed manner rather than in a centralized manner. Moreover, the rate maximization algorithm has the worst EE performance since it aims at improving the throughput of the DAS, implying that all RAUs should be activated and each RAU uses full power to transmit.

As expected, Fig.~\ref{Raterateconstraint} shows that the rates achieved by all methods increase with the rate constraint and finally keep constant at 504 Mbit/s. In this case, all RAUs should be active and full transmit power is used. When the rate constraint is beyond 504 Mbit/s, the original problem is infeasible, which can be checked by the first step of Algorithm 1. It can be seen from Fig.~\ref{Portrateconstraint} that the number of active RAUs for different RAU selection methods are almost the same and increases with the rate constraint, which is consist with the analysis for Fig. \ref{EErateconstraint}.

Fig.~\ref{EEportnum} studies the effect of the number of RAUs on the EE performance of different methods under two cases:  $R_{\rm{min}}=0\  {\rm{Mbps/Hz}}$ and $R_{\rm{min}}=800\  {\rm{Mbps/Hz}}$. The corresponding rate and number of active RAUs are shown in Fig.~\ref{Rateportnum} and Fig.~\ref{Portportnum}, respectively. Similar to the observations in Fig.~\ref{EErateconstraint}, the EE performance of the proposed RAU selection is almost the same as exhaustive search and channel-norm-based method, and significantly outperforms the other schemes for both cases. For the case of $R_{\rm{min}}=0\  {\rm{Mbps/Hz}}$, it can be observed from Fig.~\ref{EEportnum}.(a) that the EE achieved by the proposed RAU selection method increases with the number of RAUs. The reason is that as the number of RAUs increases, the average access distance of the user to the RAUs reduces, and thus the EE of {the DAS improves. On the other hand, the EE of the CAS almost stays fixed for all considered number of RAUs. This again confirms the fact that, in order to have a better EE performance, massive number of antennas in 5G networks should be placed spatially separated.  It is interesting to see from Fig.~\ref{Portportnum} that when no rate constraint is incorporated, the number of active RAUs is always equal to one. Note that a similar property has also been observed in \cite{Heejin2013} for the single-antenna case.

\begin{figure}[h]
\centering
\includegraphics[width=3.9 in]{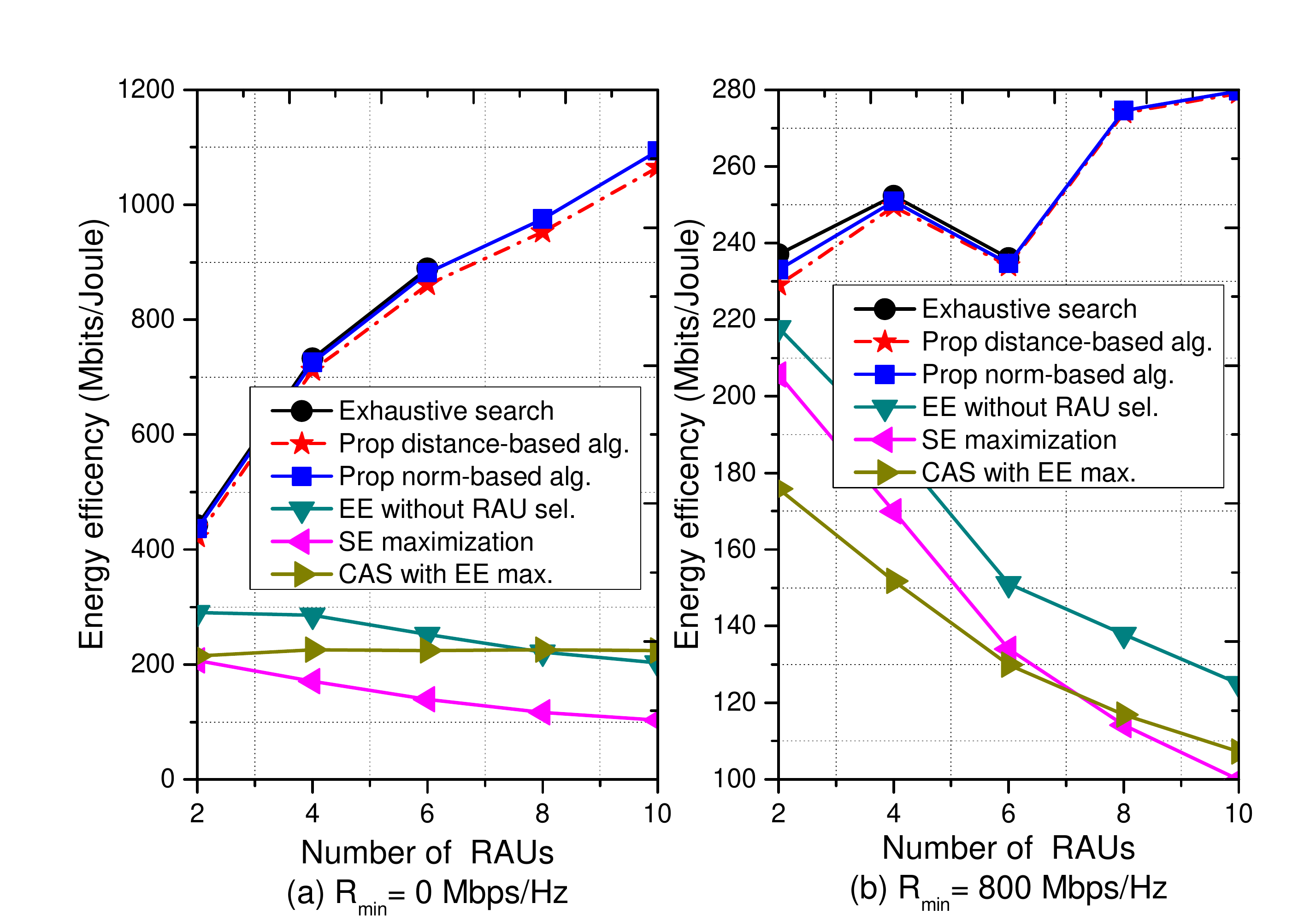}
\caption{EE under different number of RAUs with (a) $R_{\rm{min}}=0\  {\rm{Mbps/Hz}}$  and (b) $R_{\rm{min}}=800\  {\rm{Mbps/Hz}}$.}
\label{EEportnum}
\end{figure}

\begin{figure}[h]
\centering
\includegraphics[width=3.9 in]{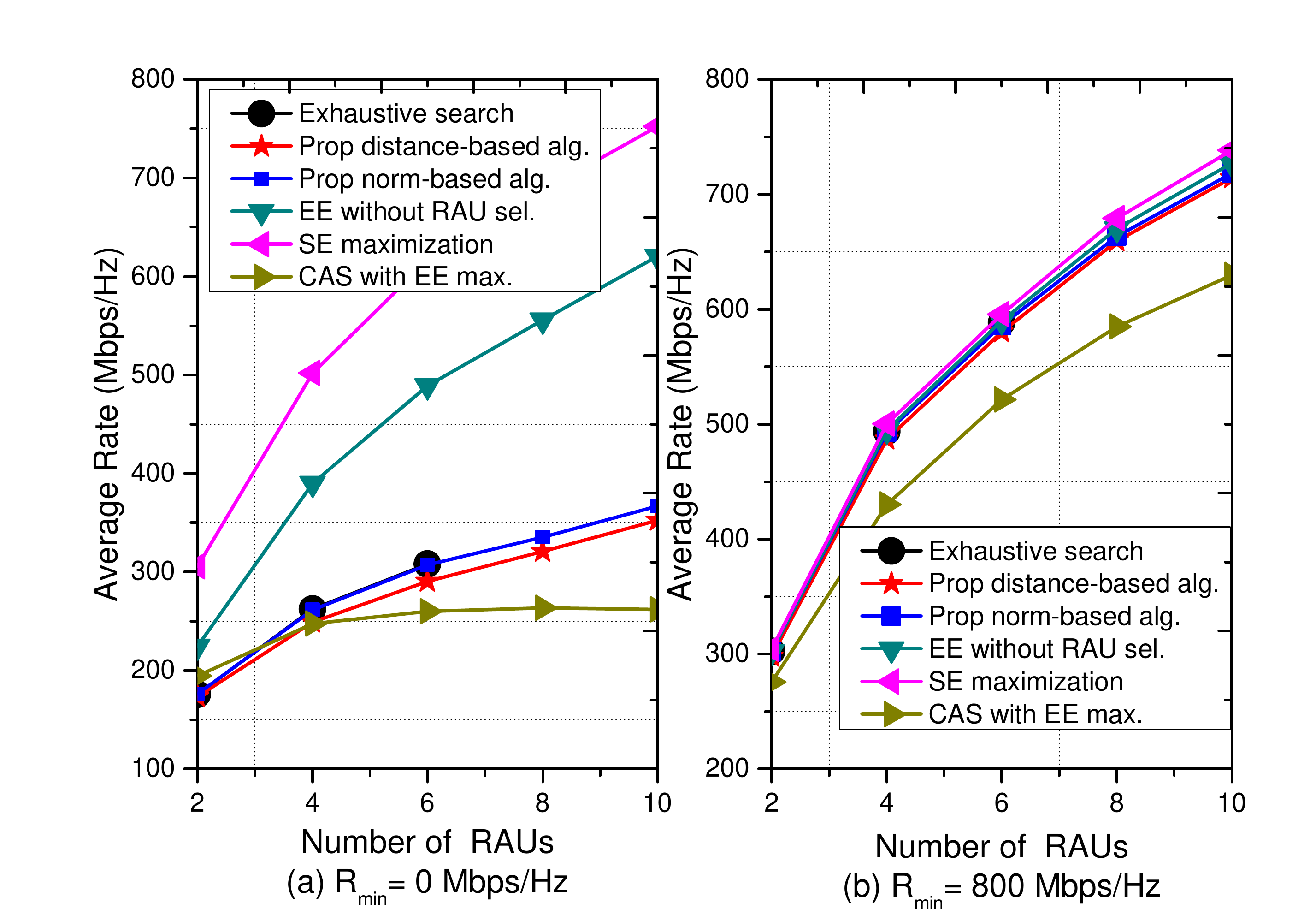}
\caption{Corresponding rate under different number of RAUs with (a) $R_{\rm{min}}=0\  {\rm{Mbps/Hz}}$  and (b) $R_{\rm{min}}=800\  {\rm{Mbps/Hz}}$.}
\label{Rateportnum}
\end{figure}

\begin{figure}[h]
\centering
\includegraphics[width=3.9 in]{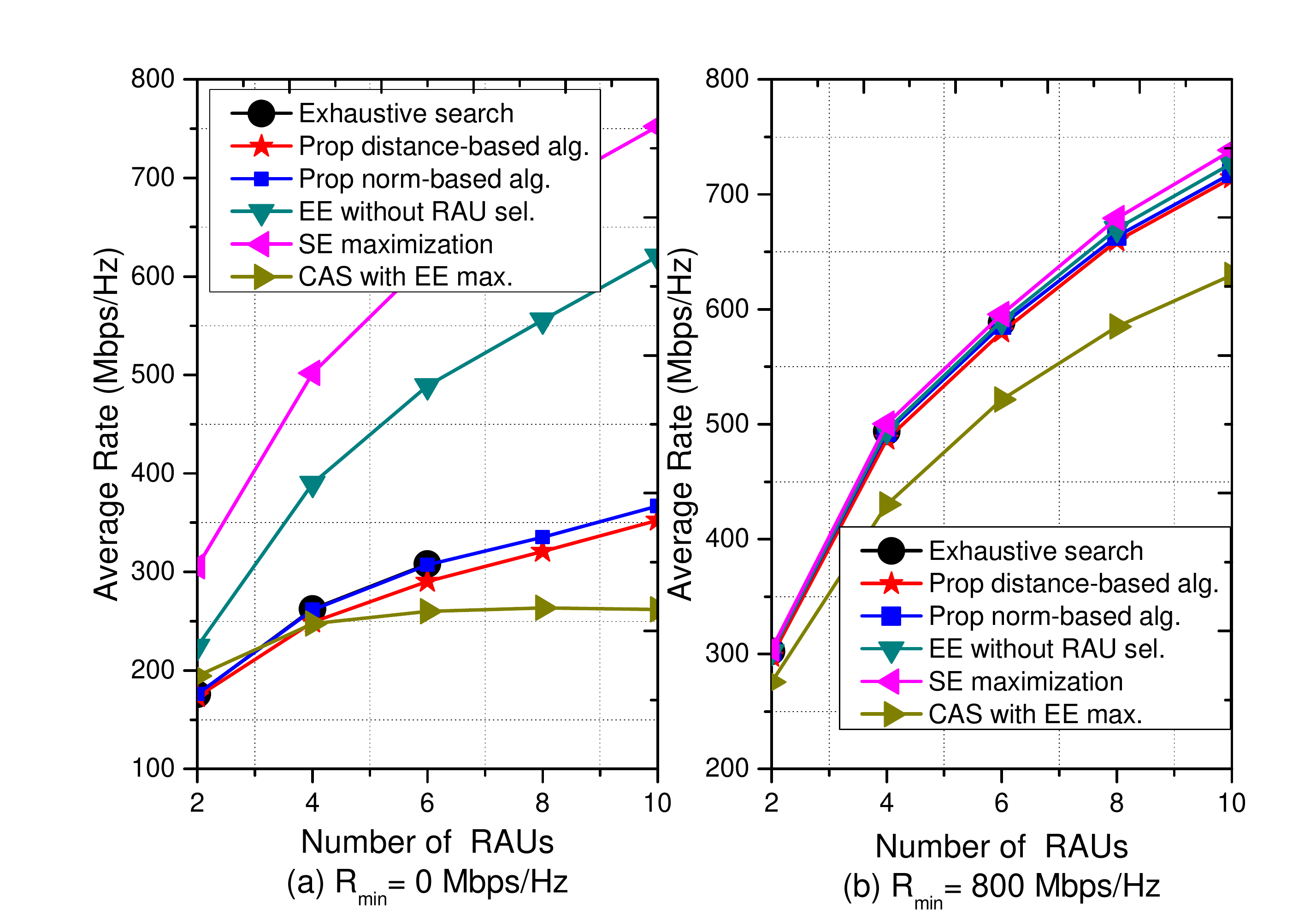}
\caption{Corresponding number of active RAUs under different number of RAUs with (a) $R_{\rm{min}}=0\  {\rm{Mbps/Hz}}$  and (b) $R_{\rm{min}}=800\  {\rm{Mbps/Hz}}$.}
\label{Portportnum}
\end{figure}

For the case of $R_{\rm{min}}=800\  {\rm{Mbps/Hz}}$, from  Fig. \ref{EEportnum} (b), we find that the EE of the CAS, i.e., the EE achieved by ``CAS with EE max'' decreases dramatically with the number of RAUs. This is due to the fact that more RAUs should be active to support the rate requirement. On the other hand, the EE achieved by our proposed RAU selection method only decreases at $I=6$. Then, with the increase of RAUs, the benefits from the RAU selection diversity dominates the negative effects of the increasing circuit power consumption.
It can also be seen from Fig.~\ref{Rateportnum} that the rate achieved by our algorithm is comparable with those achieved by the EE optimization without RAU selection and the rate maximization method. However, it has much better EE performance as seen in Fig.~\ref{EEportnum} (b). Also, Fig. \ref{Portportnum} (b) shows the number of \emph{active} RAUs increases with the number of RAUs, since more RAU should be active for the high rate requirement at $R_{\rm{min}}=800\  {\rm{Mbps/Hz}}$.

\section{Conclusion}\label{conclusion}

In this paper, we have studied the transmit covariance optimization for the EE maximization problem in a multiple-antenna DAS, where both the per-RAU power constraints and the user's rate requirement are incorporated. Given a fixed set of active RAUs, we obtain the optimal transmit covariance matrix by splitting the EE optimization problem into three subproblems, each of which has been solved with low-complexity. Then, we develop a novel distance-based RAU selection method to additionally improve the EE of the DAS with much reduced complexity. Simulation results show that our proposed RAU selection performs as well as the optimal exhaustive RAU search method, and significantly outperforms EE optimization method without RAU selection and the antenna selection method in the CAS.

\begin{appendices}

\section{Proof of Lemma 3}

Before proving the lemma, we first study the property of the following function
\begin{equation}\label{eepro}
 \text{EE}({x}) = \mathop {{\rm{max}}\;}\limits_{{\bf{Q}} \in {\cal W}} \frac{{\log _2\left| {{\bf{I}} + {\bf{HQ}}{{\bf{H}}^H}} \right|}}{{{\rm{tr}}\left( {\bf{Q}} \right) + {x}}},
\end{equation}
where $x>0$. Denote the optimal solution for a given $x$ as ${\bf{Q}}^\circ({x})$ and the corresponding achievable rate as $R(x)=\log _2\left| {{\bf{I}} + {\bf{H}}{\bf{Q}}^\circ({x}){{\bf{H}}^H}} \right|$.

\textbf{Property 1:} As $x\rightarrow+\infty$, $R(x)$ will approach the maximum achievable rate $R_{\text{max}}={\log _2}\left| {{\bf{I}} + {\bf{H}}{{\bf{Q}}_{(\text{P1})}^*}{{\bf{H}}^H}} \right|$ achieved by solving the rate maximization problem $(\rm{\textbf{P1}})$.

\textbf{Proof:} Given $x$, function $\text{EE}({x})$ can be obtained by using the Dinkelbach method as shown in Algorithm 3.
At any iteration $n$, $\eta^{(n+1)}$ is updated according to \eqref{eq28}, which will approach zero since the numerator is upper bounded by $R_{\text{max}}$ and the denominator approaches infinity. Then, at iteration $n+1$, the optimal transmit covariance matrix $Q$ is obtained by solving the problem in (\ref{fractional}), where  $\eta=\eta^{(n+1)}$ approaches zero. Hence,  the optimal solution to the problem in (\ref{fractional}) is almost equal to the solution of the rate maximization Problem $(\rm{\textbf{P1}})$, which completes the proof.\hfill $\Box$

\textbf{Property 2 \cite{p9-pan2015totally}:} Function $\text{EE}({x})$ is a continuous and strictly decreasing function of $x$, while its corresponding achievable rate $R(x)$ is a continuous and strictly increasing function of $x$.  \hfill $\Box$

Based on the above properties, we start to prove the lemma. From (\ref{eq28}), we see that  $\text{EE}({P_C})=\eta^*$. According to Algorithm 1, we are solving Problem $(\rm{\textbf{P3}})$ because the achievable  rate $R(P_C)$ is less than  $\tilde R_{\text{min}}$, i.e., $R(P_C)<\tilde R_{\text{min}}$. Moreover, since problem (\ref{EEwithfixedset}) is feasible according to Algorithm 1, we have $R(P_C)<\tilde R_{\text{min}}<R_{\text{max}}$. Then, by using Property 1 and Property 2, there must exist some $x=x'$ for which the achievable rate is equal to $\tilde R_{\text{min}}$, i.e., $R(x')=\tilde R_{\text{min}}$. Also, $x'>P_C$ means that $\eta^*=\text{EE}({P_C})>\text{EE}({x'})$ according to the strictly decreasing nature of $\text{EE}(x)$ specified in Property 2.

According to Lemma 2,  ${\bf{Q}}^\circ({x'})$ is the optimal solution to the problem in  (\ref{fractional}) with $\eta=\text{EE}({x'})$. By comparing the problem in (\ref{eq33}) with the problem in (\ref{fractional}) and recalling that $R(x')=\tilde R_{\text{min}}$, we can conclude that ${\bf{Q}}^\circ({x})$ is also the optimal solution to the problem in  (\ref{eq33}), i.e., ${\bf{Q}}^*({\mu ^*})={\bf{Q}}^\circ({x})$, and  $\mu ^*=\text{EE}({x'})$. Hence, we have $\eta^*>\mu ^*$.

\end{appendices}



\
\



\bibliographystyle{IEEEtran}
\bibliography{myre}




\end{document}